\documentclass[12pt]{article}
\usepackage{latexsym,epsfig,graphicx,amsmath,amssymb,
	amscd,multirow,chicago,psfrag,paralist,dsfont,url}
\usepackage[titletoc]{appendix}
\usepackage{booktabs}

\usepackage{epstopdf}

\newcommand{\multilinecell}[2][c]{%
	\begin{tabular}[#1]{@{}c@{}}#2\end{tabular}}
\newcommand{\thetavec}{{\boldsymbol{\theta}}}
\newcommand{\veps}{\varepsilon}
\newcommand{\vepsvec}{{\boldsymbol{\varepsilon}}}
\newcommand{\Sigmavec}{{\boldsymbol{\Sigma}}}

\newcommand{\wvec}{{\boldsymbol{w}}}
\newcommand{\yvec}{{\boldsymbol{y}}}
\newcommand{\zvec}{{\boldsymbol{z}}}
\newcommand{\zerovec}{{\boldsymbol{0}}}
\newcommand{\onevec}{{\boldsymbol{1}}}

\newcommand{\Indfun}{{\mathds{1}}}
\newcommand{\betavec}{{\boldsymbol{\beta}}}

\newcommand{\E}{\mathbf{E}}

\newcommand{\muhat}{\widehat{\mu}}
\newcommand{\sigmahat}{\widehat{\sigma}}

\newcommand{\wh}{\widehat}

\newcommand{\xvec}{\boldsymbol{x}}

\newcommand{\nuvec}{{\boldsymbol{\nu}}}
\newcommand{\rhovec}{{\boldsymbol{\rho}}}
\newcommand{\muvec}{{\boldsymbol{\mu}}}

\newcommand{\tr}{{\rm tr}}
\newcommand{\corr}{{\rm Corr}}

\newcommand{\loglik}{\mathcal{L}}

\newcommand{\zmat}{\mathbf{Z}}
\newcommand{\xmat}{\mathbf{X}}
\newcommand{\cmat}{\mathbf{C}}
\newcommand{\umat}{\mathbf{U}}
\newcommand{\vmat}{\mathbf{V}}
\newcommand{\wmat}{\mathbf{W}}

\newcommand{\Bmat}{\mathbf{B}}
\newcommand{\Amat}{\mathbf{A}}

\newcommand{\Pmat}{\mathbf{P}}
\newcommand{\Lmat}{\mathbf{L}}

\newcommand{\epsilonvec}{\boldsymbol{\varepsilon}}
\newcommand{\sigmamat}{\mathbf{\Sigma}}
\newcommand{\omegamat}{\mathbf{\Omega}}

\newcommand{\alphavec}{\boldsymbol{\alpha}}

\newcommand{\lambdamat}{\mathbf{\Lambda}}

\def\T{{ \mathrm{\scriptscriptstyle T} }}

\begin{document}
	
	\title{Prediction for Distributional Outcomes in High-Performance Computing I/O Variability}
	
	
	\author{Li Xu$^1$, Yili Hong$^2$, Max D. Morris$^3$, and Kirk W. Cameron$^4$\\
		\small $^1$Department of Epidemiology, Harvard University, Boston, MA, 02215\\
		\small $^2$Department of Statistics, Virginia Tech, Blacksburg, VA 24061\\
		\small $^3$Department of Statistics, Iowa State University, Ames, IA 50011\\
		\small $^4$Department of Computer Science, Virginia Tech, Blacksburg, VA 24061\\
	}

\date{}	

	\maketitle

	\begin{abstract}
		Although high-performance computing (HPC) systems have been scaled to meet the exponentially-growing demand for scientific computing, HPC performance variability remains a major challenge and has become a critical research topic in computer science. Statistically, performance variability can be characterized by a distribution. Predicting performance variability is a critical step in HPC performance variability management and is nontrivial because one needs to predict a distribution function based on system factors. In this paper, we propose a new framework to predict performance distributions. The proposed model is a modified Gaussian process that can predict the distribution function of the input/output (I/O) throughput under a specific HPC system configuration. We also impose a monotonic constraint so that the predicted function is nondecreasing, which is a property of the cumulative distribution function. Additionally, the proposed model can incorporate both quantitative and qualitative input variables. We evaluate the performance of the proposed method by using the IOzone variability data based on various prediction tasks. Results show that the proposed method can generate accurate predictions, and outperform existing methods. We also show how the predicted functional output can be used to generate predictions for a scalar summary of the performance distribution, such as the mean, standard deviation, and quantiles. Our methods can be further used as a surrogate model for HPC system variability monitoring and optimization.

	\end{abstract}

	\section{Introduction}

\begin{figure}
		\centering
\begin{tabular}{cc}
\includegraphics[width=0.47\textwidth]{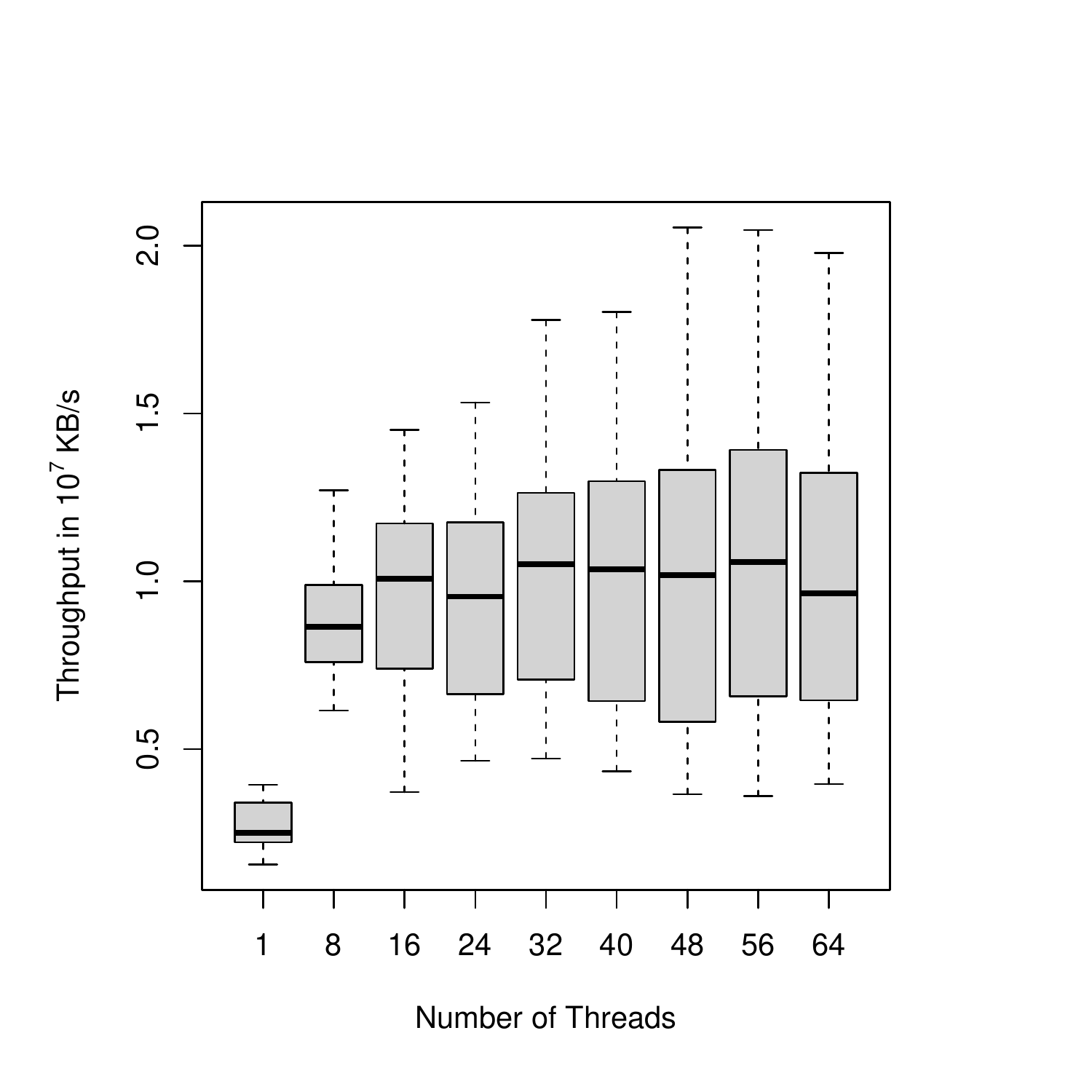}&
\includegraphics[width=0.47\textwidth]{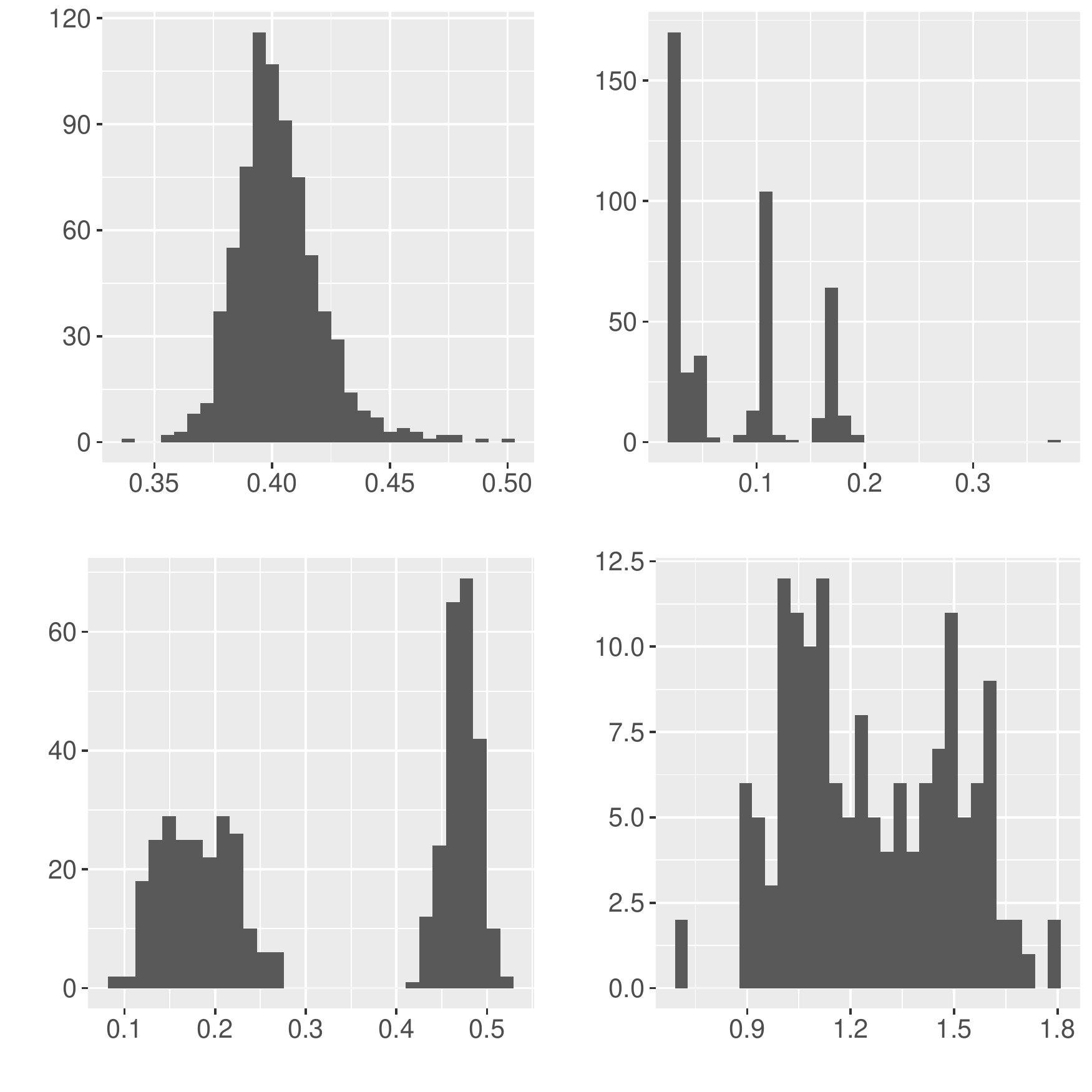}\\
(a) Example of I/O Variability & (b) Throughput Distributions
\end{tabular}
\caption{(a) Example of I/O variability in throughput as a function of the number of threads in the IOzone data, and (b) histograms for the I/O throughput under four specific HPC system configurations showing examples of distributions with various shapes. The $x$-axis is the throughput ($10^7$ KB/s) and the $y$-axis is the frequency.}\label{fig:ch4hist4}
	\end{figure}

	High-performance computing (HPC) systems aggregate a large number of computers to provide a high level of computing performance. In the past decades, the performance of HPC systems has been increased to meet the exponentially-growing demand for scientific computing. However, existing work (\shortciteNP{rahimi2015task};  \shortciteNP{Cameron-MOANA-2019}) has observed that the performance variability increases with HPC system scale and complexity. For example, Figure~\ref{fig:ch4hist4}(a) shows that the input/output (I/O) throughput, as one measure of the system performance, increases as the number of threads increases based on a subset of the IOzone data to be introduced in Section~\ref{sec:iodata}. However, we observe that the performance variability, as shown by the boxplots, also increases. Existing studies reveal that variability can influence the performance in many aspects from hardware (\shortciteNP{6493621}), middleware (\shortciteNP{akkan2012stepping}; \shortciteNP{ouyang2015achieving}) to applications (\shortciteNP{hammouda2015noise}). Thus performance variability management has become an important research area in computer science, which is affected by system configurations (e.g., CPU frequency). Unfortunately, the quantitative relationship between the system configuration and variability is not clear, which makes the HPC performance variability management challenging. Studies have discovered that the relationship between HPC variability and system configuration is complicated (\shortciteNP{lux2018novel}; \shortciteNP{chang2018predicting}). To study the complicated relationship, statistical tools can be useful for data collection, model building, and performance variability prediction. Large-scale experiments are essential to provide sufficient data for modeling the complex variability map, and experimental design tools have been used for efficient data collection (\shortciteNP{wang2020JQT}).
	
Regarding modeling and prediction, performance variability can be characterized by a distribution.  Most existing work in computer science, however, only uses a summary statistic to represent the level of variability. For example, \shortciteN{Cameron-MOANA-2019} study the standard deviation of the IOzone throughput. \shortciteN{xu2020modeling} show that the throughput distribution is multimodal so a summary statistic like standard deviation cannot represent the system variability. As an illustration, Figure~\ref{fig:ch4hist4}(b) shows the histograms of the I/O throughput under four specific HPC system configurations. The top left panel shows a distribution with one mode, and the bottom left panel shows a mixture of two components, while the right two panels show a mixture of three and more than three components. Therefore, the distributions of the throughput are complicated, and it is typically not sufficient to use summary statistics or a simple parametric distribution to describe them.
	
Because the performance distribution is complicated, it will be ideal to have a general method to predict the entire distribution. Furthermore, various metrics are often of interest in the HPC study. The mean or median of the throughput distribution can be used as an overall performance measure, while the standard deviation can be used as a measure of variability or stability. Various quantiles of the performance distribution can serve as practical lower or upper bounds of throughput, which leads to a general need for modeling and predicting the performance distribution. This is because once the distribution is predicted, one can derive all the above-mentioned metrics, which brings tremendous benefits in HPC variability management.

	To address the challenging problem in HPC variability management, the main objective of this paper is to generate distributional-output predictions for HPC variability study. The prediction framework is outlined as follows. We first use I-splines to smooth the discrete sample quantile function and the obtained spline coefficients matrix is then used to represent the distribution function. Singular value decomposition (SVD) is implemented to reduce the dimension of the coefficient matrix. For prediction, we propose a special Gaussian process (GP) named linear mixed Gaussian process (LMGP), that incorporates both quantitative and qualitative variables. The expectation-maximization (EM) algorithm is used to estimate the parameters. Results show that our prediction framework can achieve accurate predictions under HPC setting. To the best of our knowledge, this work is the first work that develops a statistical framework predicting the distributional outcome with mixed types of inputs and modeling HPC throughput distributions along with their associated measures of variability.
	
	We give a brief literature review on computer experiments with an emphasis on mixed types of input and output. Computer experiments are often constructed to emulate a physical system. Due to the complexity and expense of evaluating system behavior, a surrogate model is usually used to describe the system behavior based on the data collected by the experiments. Popular surrogate models include response surfaces (\shortciteNP{Box1951rsm}), Gaussian process models (\shortciteNP{10.5555/1162254}), localized linear regression (\shortciteNP{doi:10.1080/01621459.1979.10481038}), and their extensions. \shortciteN{tgpjasa}, \shortciteN{Chipman2002}, \shortciteN{bartanas}, and \shortciteN{dynatree} use the binary tree to divide the input space and fit separate Gaussian process in each sub-region. Multivariate adaptive regression splines (MARS) uses splines and stepwise regression to model the complex relationships between input and output (\shortciteNP{friedman1991}). To determine the best model with respect to node location and number of nodes, a generalized cross-validation procedure (\shortciteNP{hastie2009elements}) is used to do model selection. The linear Shepard (LSP) algorithm uses radial basis functions to design weight and build a localized linear regression model (\shortciteNP{Thacker}).

	While most of those models assume the inputs of surrogates are continuous, categorical inputs are common in application. For example, in the HPC setting, the type of storage has two options: solid-state drive (SSD) and hard disk drives (HDD). To utilize categorical variables, \shortciteN{zhou2011simple} propose the CGP and \shortciteN{doi:10.1080/00401706.2016.1211554} extend the CGP with an additive model structure. In addition, most existing methods focus on scalar prediction, while the output of some engineering models can be complicated (\shortciteNP{bayarri2007}). Examples of applications with complicated outputs include the boundary condition of a partial differential equation (\shortciteNP{doi:10.1080/00401706.2017.1345702}), the thermal-hydraulic computations (\shortciteNP{AUDER2012122}), and the satellite orbiting carbon observatory (\shortciteNP{ma2020computer}).

	For the work on computer experiments modeling with functional outputs, \shortciteN{doi:10.1080/00401706.2013.869263} develop a Monte Carlo expectation-maximization (MCEM) algorithm to convert the irregularly spaced data into a regular grid so that the Kronecker product-based approach can be employed for efficiently fitting a kriging model to the functional data. \shortciteN{higdon2008computer} provide a dimension-reduction method to the high-dimensional output computer experiments. \shortciteN{FanJiang2020} provide a robust parameter design to computer models with multiple functional outputs. \shortciteN{FRUTH2015260} conduct sensitivity analysis method for functional input. \shortciteN{Dorin2010funa} proposes a framework called functional ANOVA to analyze the computer experiments with time series outputs. However, to our best knowledge, there is no work focused on the distributional outcome on computer models with both qualitative and quantitative inputs, which cannot be addressed by straightforward applications of existing methods.
	
	Because of the distributional outcome, the properties of the distribution functions need to be met. Specifically, the cumulative distribution function is right-continuous and nondecreasing. In addition, effective modeling of output distributions generally requires large datasets because complicated experiments are essential to capture the distributional information. Given the need to predict the distribution and the fact that the distribution is complicated, we use the Gaussian process models as the basis for our work. Compared to parametric models, the Gaussian process can establish a more complicated relationship between the input and response variables. In this paper, we propose a prediction framework with Gaussian process that can predict the distributional output given both quantitative and qualitative inputs.

	The rest of this paper is organized as follows. Section~\ref{sec:iodata} describes the HPC IOzone data. Section~\ref{sec:mod} describes the prediction framework including the curve representation, the formulation of the LMGP model, the EM algorithm for parameter estimation, and the functional prediction. Section~\ref{sec:res} presents the prediction results on the IOzone data for different input and output (I/O) operation modes in predicting the quantile functions. Section~\ref{sec:sumstat} shows the comparison results with those existing models in predicting summary statistics of the throughputs. Section~\ref{sec:dis} discusses the results and several areas for future work.
	
	\section{HPC Performance Study}\label{sec:iodata}

	While the system variability has many aspects, we concentrate on the I/O tasks as these types of the procedure will reveal the highest variability and exhibit the most interesting system performance characteristics. I/O is identified as a high variation operation and the IOzone benchmark (\shortciteNP{capps2008iozone}) is used to collect performance data on the various system I/O operations. The reported throughput values are used to represent the system performance and furthermore, the variation of the throughput under identical system configurations is treated as the system variability. The unit of the throughput is KB/s. For convenience, all the throughputs in this paper are on the scale of $10^7$ KB/s.
	
\begin{figure}
		\centering
		\includegraphics[width=0.6\textwidth,angle=270,origin=c]{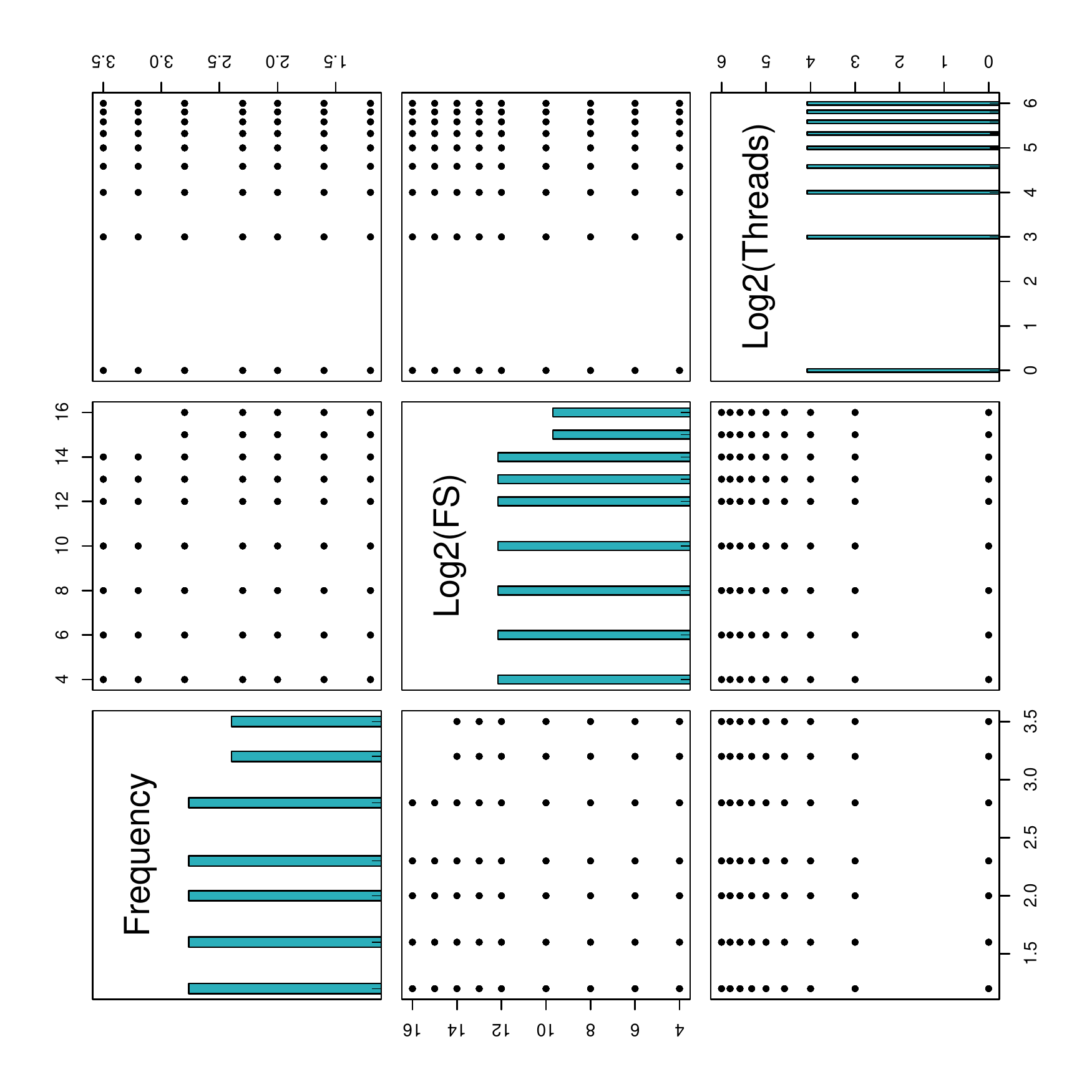}
		\caption{Scatter plots and histograms for three continuous factors under I/O operation mode initial\_writer. ``Log2(FS)" means the binary logarithm of the file size. The unit for CPU Frequency is GHz and the unit for file size is KB.}\label{fig:fda_inidist}
	\end{figure}

	The configurations are characterized by a list of variables, which are referred to as inputs. There are two kinds of inputs, namely, numerical inputs and categorical inputs. We have four numerical inputs, the file size, the record size, the CPU frequency, and the number of threads. The record size is fixed at 16 KB throughout the whole experiment. Thus, the numerical variables we model in this paper are file size, CPU frequency, and the number of threads. The categorical input is the I/O operation mode, which has six levels. There are various combinations of those three continuous inputs under each level of the categorical input (i.e., the I/O operation mode). Table~\ref{tab:iozone} shows the system configurations and all possible levels we have considered in our data collecting experiments. In total, we have 22{,}734 combinations (system configurations) in the IOzone database. Figure~\ref{fig:fda_inidist} shows the combinations of continuous inputs under I/O operation mode initial\_writer. Because the levels of the file size and the number of threads are spaced on an exponential scale, we take the binary logarithm of the two variables in our subsequent analyses.

The configurations are denoted by $\{\xvec_i, \zvec_i\}, i=1,\ldots, n$. Here, $\xvec_i=(x_{i1}, \ldots, x_{ip})^{\T}$ is a $p\times 1$ vector that denotes the numerical inputs, $\zvec_i$ is a $q\times 1$ vector that denotes the categorical inputs, and $n$ is the number of configurations. In the IOzone data, $p=3$ and $q=1$. We denote the numerical input matrix by $\xmat=(\xvec_1, \ldots, \xvec_n)^{\T}$, which is of size $n\times p$, and denote the categorical input matrix by $\zmat=(\zvec_1, \ldots, \zvec_n)^{\T}$, which is of size $n\times q$. Thus the input is represented by $\{\xmat,\zmat\}$.
	\begin{table}
		\centering
		\caption{System factors and their levels used in the study of I/O variability.}\label{tab:iozone}
		\begin{center}
			\begin{tabular}{ccc}\hline
				\multilinecell{System \\ Parameters} & \multilinecell{No. of \\Levels} & Levels\\ \hline
				\multilinecell{CPU Clock \\Frequency (GHz)} & 7 & \multilinecell{1.2,  1.6, 2.0, 2.3, 2.8, 3.2, 3.5}\\\hline
				\multilinecell{Number of Threads} & 9 & \multilinecell{1, 8, 16, 24, 32, 40, 48, 56, 64}\\\hline
				\multilinecell{File Size (KB)} & 10 & \multilinecell{4, 16, 64, 256, 1024, 4096,\\ 8192, 32768, 65536}\\\hline
				\multilinecell{I/O Operation \\ Mode} & 6 & \multilinecell{random\_reader, initial\_writer,\\ random\_writer, rereader, reader, rewriter}\\\hline
			\end{tabular}
		\end{center}
	\end{table}
	
	To collect the data which reflects the distributional information, we fix the system configuration at a given combination in Table \ref{tab:iozone} and run the IOzone benchmark for a specified number of replicates. The output of each experiment run is the throughput of IOzone, which measures the I/O speed under the current system configuration. The throughput data at configuration $i$ are denoted by $y_{ib}$, $i=1, \ldots, n$, and $b=1,\ldots, m_i$. Let $\yvec_i=(y_{i1}, \ldots, y_{im_i})^{\T}$, where $m_i$ is the number of replicates for $i$th configuration. The values of $m_i$ vary from 150 to 900, depending on the specific system configuration. It took several months to collect all the data over a Linux server. In particular, the experiments were conducted on a 12-node server and all the nodes are identical Dell PowerEdge R630s. Each node is equipped with Intel(R) Xeon(R) CPU E5-2637 v4@3.50 GHz, 16 GB DRAM (2 DIMMs), and a new 200 GB SSD with Intel model SSDSC2BA200G4R. There are 2 sockets with 4 cores per socket. In total, there are 8 physical cores and 16 CPUs with hyper-threading enabled. The operating system is Debian GNU/Linux with kernel version 4.14 and the IOzone version is 3.465. Note that we are working with real performance data from HPC systems (not with emulator data as in some computer experiment literature).

The data $\yvec_i$ are then used to estimate a distribution function which we will treat as a functional response in modeling. For the development of the model and notational convenience, we need to first sort the data by $\zmat$. Let $c$ be the number of unique combinations of the categorical variables (i.e., the unique rows in $\zmat$). We sort the data $\{\yvec_i,\xvec_i,\zvec_i\},i=1,\ldots,n,$ by the unique categorical combinations. Let $n_k$ be the number of rows in the $k$th categorical combination in $\zmat$ for $k=1,\ldots,c$.

	\section{The Prediction Framework}\label{sec:mod}
	Our proposed framework for predicting HPC throughput distribution has three components: curve representation, Gaussian process for prediction, and reconstruction of functional curves.
	\subsection{Curve Representation}
	For the throughput data, $\yvec_i$, from configuration $i$, we are interested in its cumulative distribution function (CDF), $F_i(y)$. Because the distribution of the throughput is usually complicated and cannot be adequately described by commonly used parametric distributions, we use the empirical cumulative distribution function (ECDF) to estimate the distribution function. In particular, the ECDF is computed as
	$$
	\wh{F}_i(y)=m_i^{-1}\displaystyle\sum_{b=1}^{m_i}\Indfun(y_{ib}\leq y).
	$$
	The critical points in the ECDF are $\{[y_{i(b)}, b/m_i], b=1,\ldots, m_i\}$, where $y_{i(b)}$ is the sorted version of $y_{ib}$ in the ascending order.
	
\begin{figure}
		\begin{center}
			\begin{tabular}{cc}
				\includegraphics[width=.47\textwidth,angle=270,origin=c]{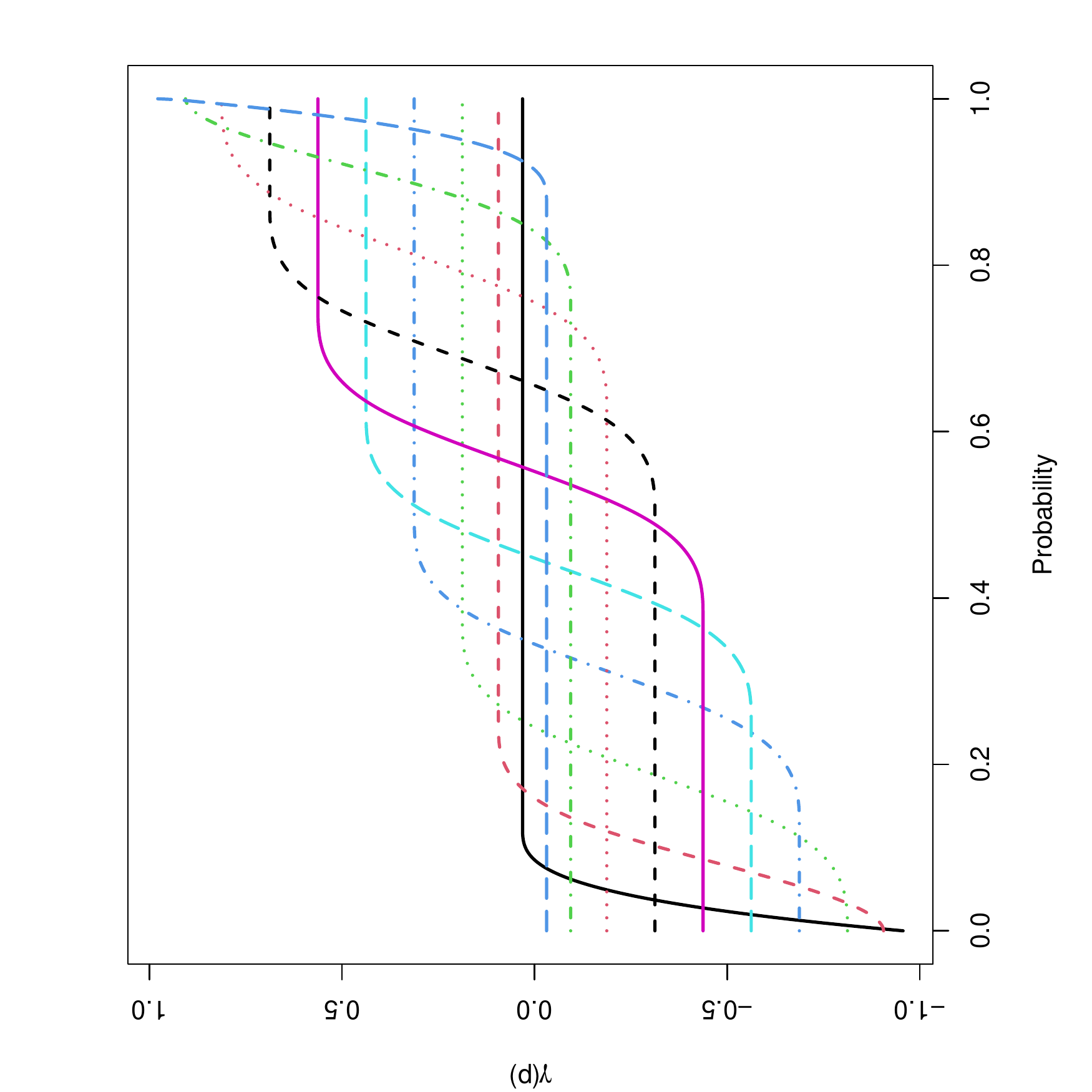}&
				\includegraphics[width=.47\textwidth,angle=270,origin=c]{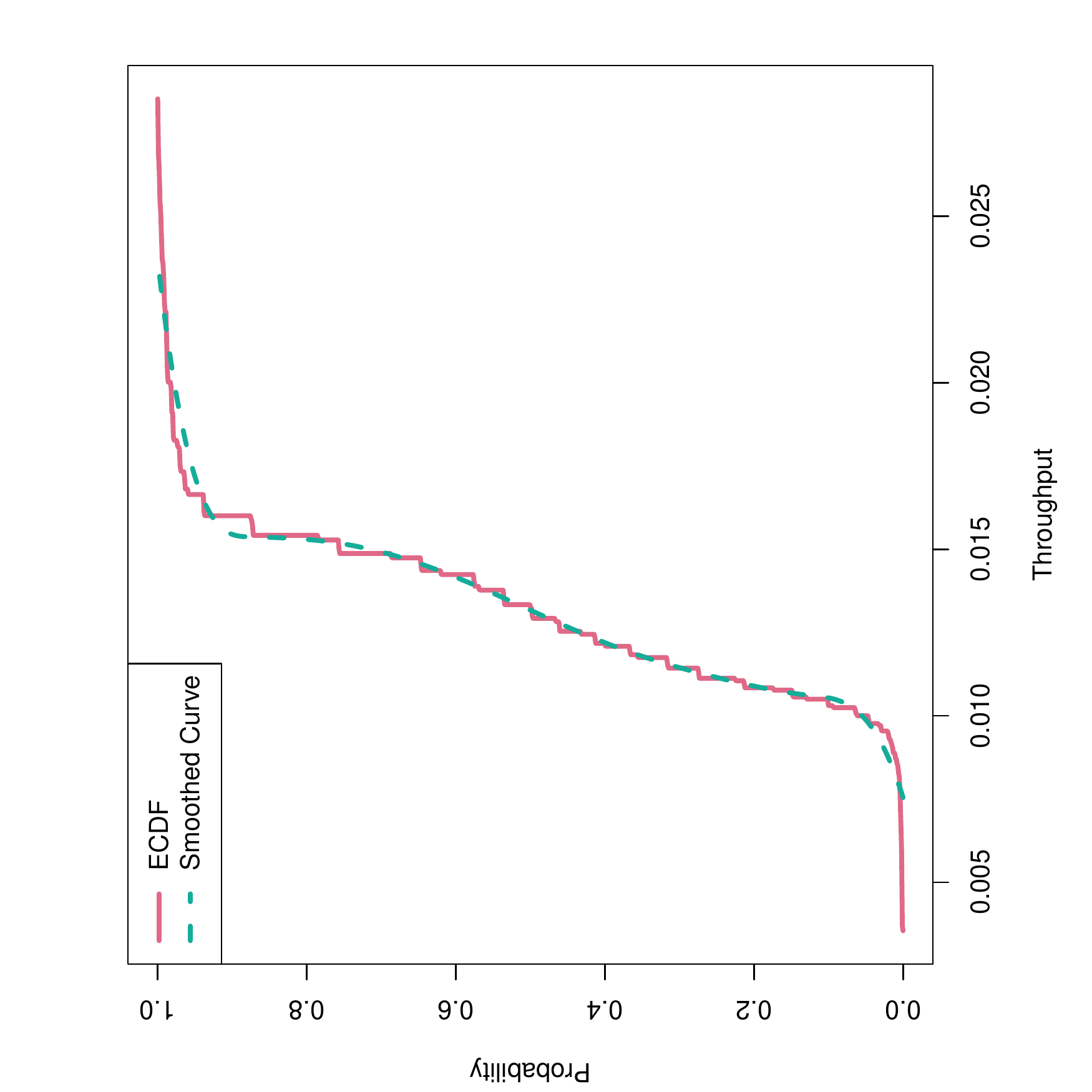}\\
				(a) I-spline Bases & (b) ECDF
			\end{tabular}
		\end{center}
		\caption{Examples of I-spline bases (a), and plots of the ECDF as a step function and its corresponding smoothed curve for a specific system configuration (b).}\label{fig:ispline}
	\end{figure}

\begin{figure}
		\begin{center}
			\begin{tabular}{ccc}
				\includegraphics[width=.31\textwidth]{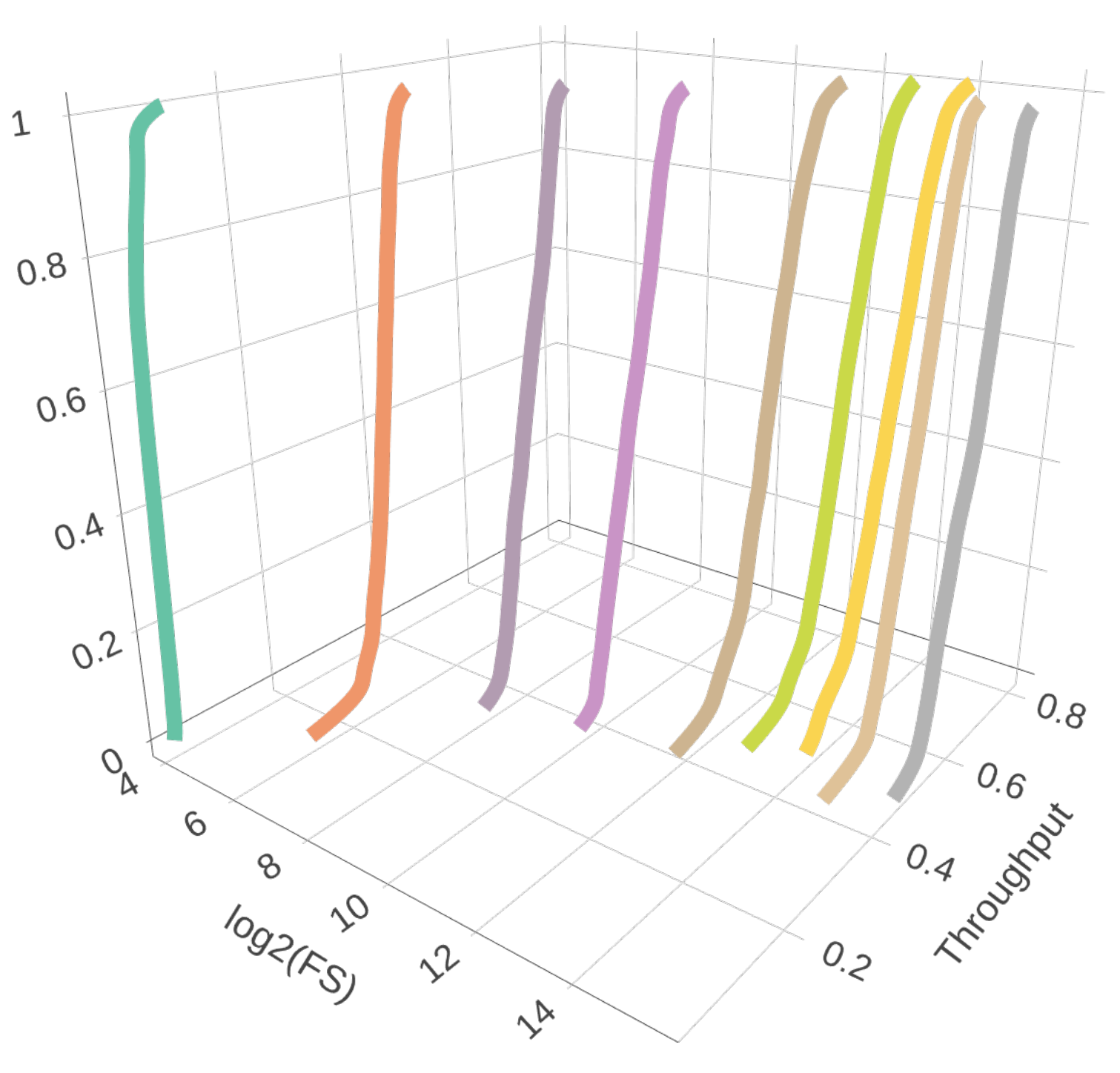}&
				\includegraphics[width=.31\textwidth]{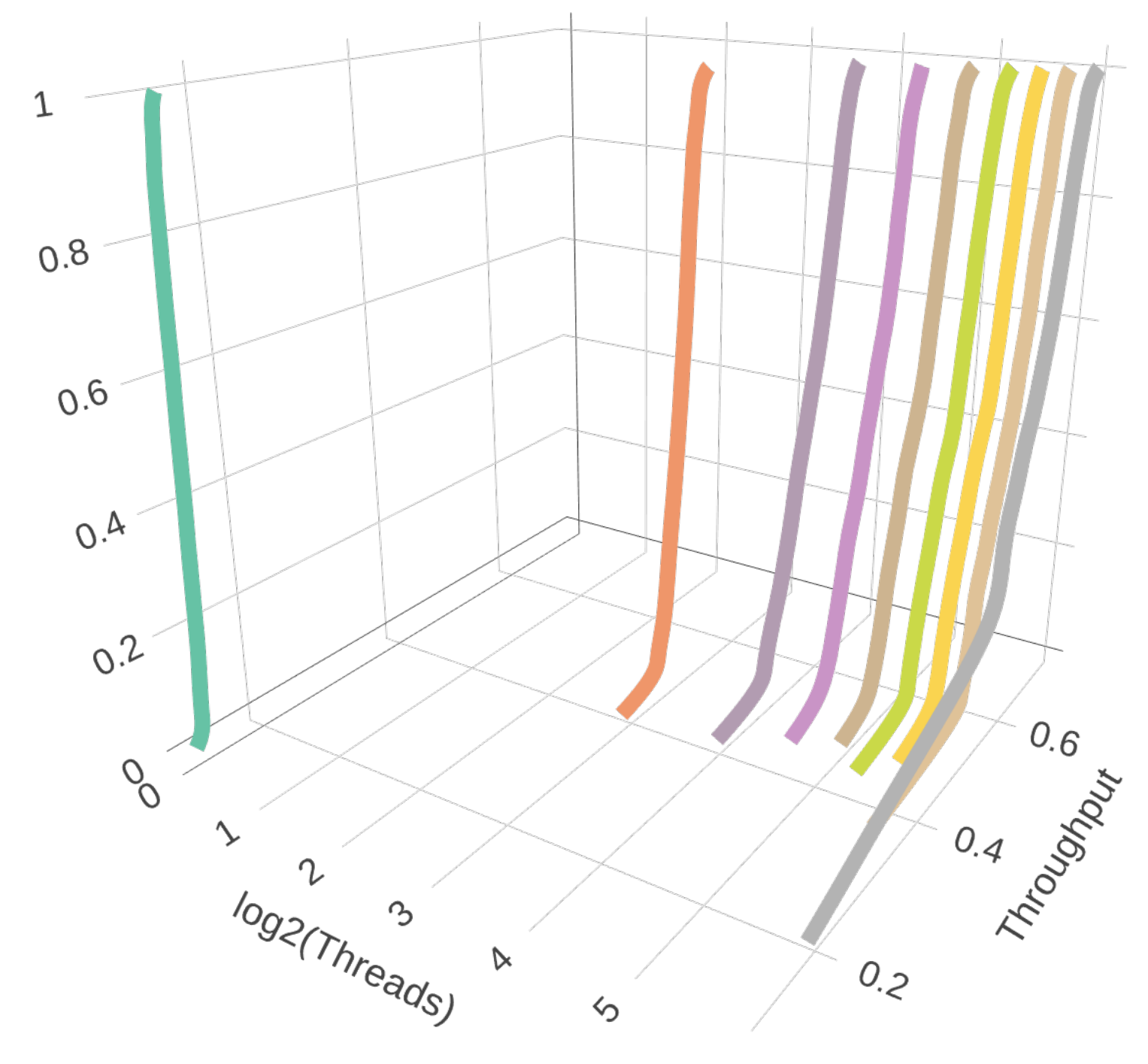} &
				\includegraphics[width=.31\textwidth]{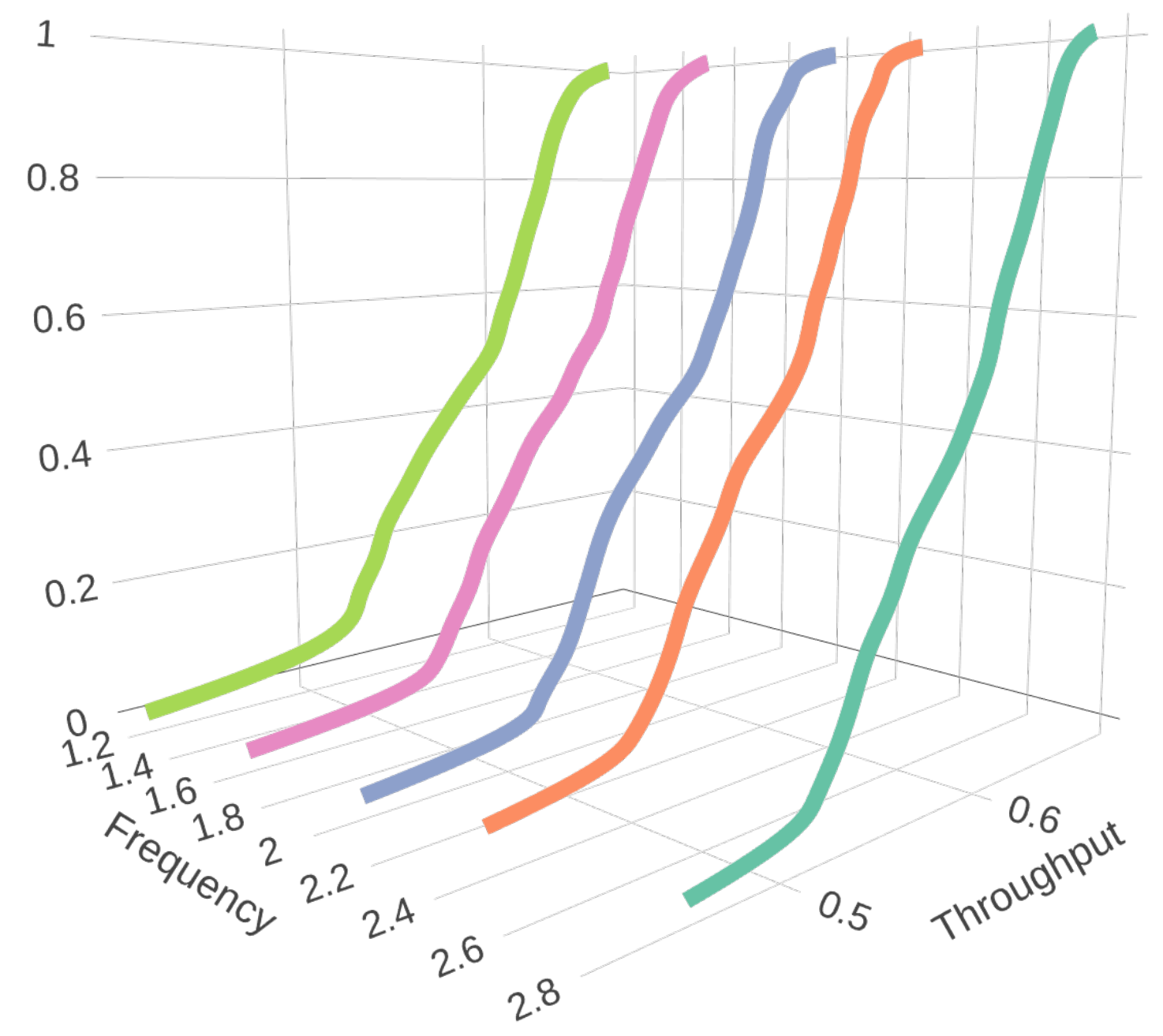}\\
				(a) File Size & (b) Threads& (c) Frequency\\
			\end{tabular}
		\end{center}
		\caption{Smoothed CDFs when one continuous configuration changes. The $z$-axis is the probability. In (a), file size changes while (frequency, threads) is fixed at $(2.3,32)$. In (b), the number of threads changes while (file size, frequency) is fixed at $(65536,2.3)$. In (c), frequency changes while (file size, threads) is fixed at $(65536,32)$.}\label{fig:3Dprob}
	\end{figure}

	Because the ECDF is only right continuous and always has jump discontinuities, for the convenience of modeling, we use a smooth function to approximate it. In addition, because the CDF is a non-decreasing function, we use monotonic splines for smoothing. In particular, we use I-splines (\shortciteNP{ramsay1988monotone}). I-splines are a set of functions that are positive and monotone increasing in a closed interval and constants outside this closed interval. Figure~\ref{fig:ispline}(a) shows the curve of a set of I-spline basis functions. Figure~\ref{fig:ispline}(b) shows one example of the ECDF and the curve after being smoothed. Figure~\ref{fig:3Dprob} shows how the smoothed CDF curve changes when we vary on one of the three continuous configuration factors. We find a complicated relationship between the CDF curves and input configurations. For example, in Figure~\ref{fig:3Dprob}(c), when we have more threads, the range of the throughputs will have a larger range and the shape of the curve also changes. Figure~\ref{fig:3Dprob} indicates that predicting the distributional outcome is challenging.
	
	
\begin{figure}
		\begin{center}
			\begin{tabular}{cc}
				\includegraphics[width=.45\textwidth,angle=270,origin=c]{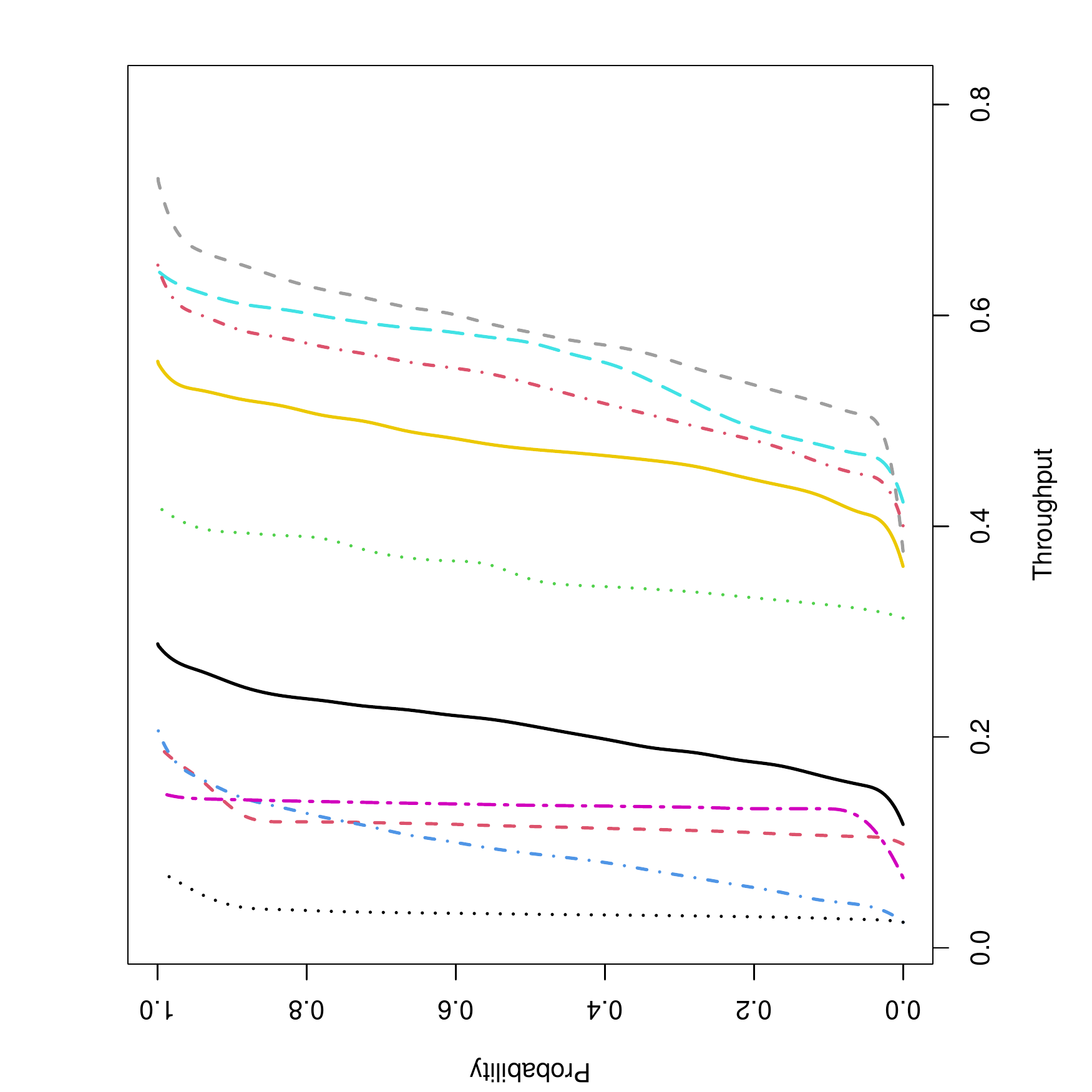} &
				\includegraphics[width=.45\textwidth,angle=270,origin=c]{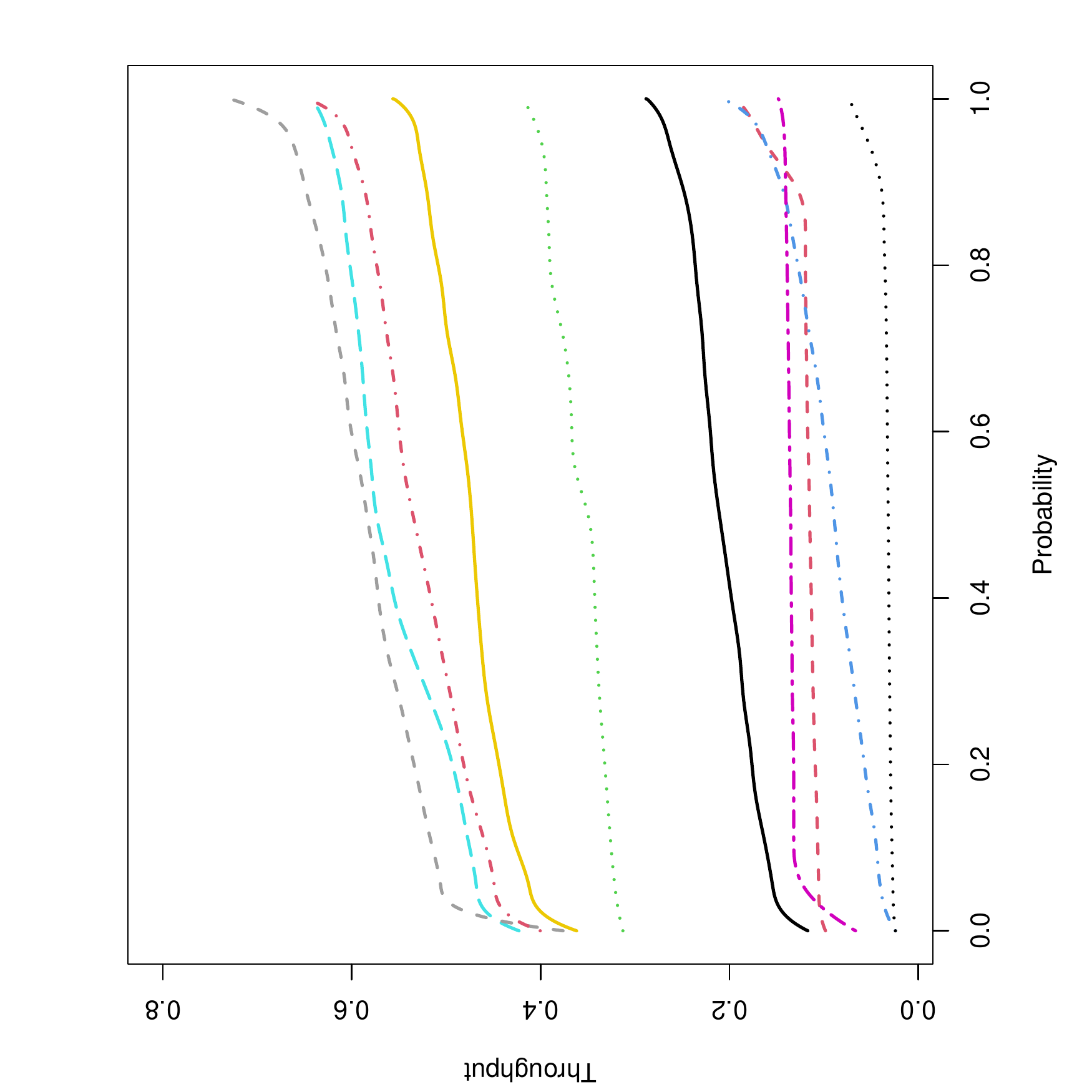} \\
				(a) CDF & (b) Quantile Function
			\end{tabular}
		\end{center}
		\caption{Examples of I-spline smoothed CDF and quantile function for the throughputs under 10 randomly picked configurations.}\label{fig:cdf}
	\end{figure}

	A set of knots are needed to construct the spline bases. Based on an initial exploration of the data, we find that the supports of the CDF are quite different for different configurations. Figure~\ref{fig:cdf}(a) shows a typical example in the IOzone data that the distributions of the throughput from different configurations have different supports. The supports of each CDF vary largely, which is challenging to choose both the number of and the locations of the spline knots. In order to cover the entire range of the CDF support and ensure smoothing accuracy, we would need a large number of knots. To overcome this difficulty, we need to set our predicted probability function to have common support with a fixed boundary. We show ten examples of smoothed quantile functions in Figure~\ref{fig:cdf}(a). Each CDF is smoothed individually by a unique set of I-splines. The number of knots is 20 and the range of knots is equal to the range of throughputs under this configuration. From the figure, we can see that the supports of different smoothed CDFs are different. To set common bases for all configurations, we smooth the estimated quantile function, instead of the ECDF. Because a quantile function, $Q(p)$, is defined in $(0,1)$ which is bounded, one can easily set the knots in the bounded domain.

To summarize the idea, we want to model and predict CDF's, and use spline fits to represent them.
However, splines require knot locations, which is impractical here because the support and complexity of CDF varies substantially for different experimental conditions. Thus, we instead directly model the inverse of the CDF, the quantile function, which is always defined on the same interval.

Specifically, we first construct $(d-1)$ common spline bases, $\gamma_{j}(p), j=1, \ldots, (d-1)$ and use these base to smooth the points $\{[b/m_i, y_{i(b)}], b=1,\ldots, m_i\}$ separately for configuration~$i$. Let $\betavec_i=(\beta_{i0}, \beta_{i1}, \ldots, \beta_{i, d-1})^{\T}$ be the coefficients of the spline fitting for configuration~$i$. The first element $\beta_{i0}$ is an intercept term, and the rest elements $(\beta_{i1}, \ldots, \beta_{i, d-1})^{\T}$ are the coefficients for those $(d-1)$ spline basis functions. Thus $\betavec_i$ is of size $d\times 1$. The smoothed quantile function is
	$$
	\wh Q(p)=\widehat{\beta}_{i0}+\sum_{j=1}^{d-1}\widehat{\beta}_{ij}\gamma_j(p).
	$$
	For I-splines, the intercept $\beta_{i0}$ is unconstrained, and the spline coefficients $(\beta_{i1}, \ldots, \beta_{i, d-1})^{\T}$ are constrained to be nonnegative, to ensure monotonicity. The constrained least-squares method (e.g., \citeNP{Meyer2008}) is used to find the spline coefficients. Let $\Bmat=(\betavec_1^{\T}, \ldots, \betavec_n^{\T})^{\T}$ be the spline coefficient matrix, which is of size $n\times d$. Using the spline representation, the distributional data are now represented by the coefficient matrix $\Bmat$.
	
	We apply the singular value decomposition (SVD) to de-correlate the columns of $\Bmat$, which is similar to the treatment in \shortciteN{higdon2008computer}. That is, we express $\Bmat$ as
	$$
	\Bmat=\umat\lambdamat \vmat^\T,
	$$
	where $\umat$ is an $n\times n$ unitary matrix, $\lambdamat$ is an $n\times d$ diagonal matrix with diagonal elements $\{\lambda_1,\ldots,\lambda_d\}$ which are the singular values, and $\vmat$ is a $d \times d$ unitary matrix in the SVD. Let
	\begin{align}
		\wmat=\umat\lambdamat=\Bmat\vmat.
	\end{align}
	Note here $\wmat$ is an $n\times d$ matrix. Let $\wvec_j$ be $j$th column of $\wmat$ and $w_{ij}$ be the $(i,j)$th element of $\wmat$. After re-expression of the $\Bmat$ matrix using the SVD, we focus on the resulting $\wmat$ matrix. We perform separate modeling for $\wvec_j$ because the $\wvec_j$'s are linear independent components.
	
	\subsection{Gaussian Process Modeling}
	We first summarize the formulas for modeling and prediction with the Gaussian process for continuous scalar output $\yvec$ and continuous covariates $\xmat$. Then, with the overall mean $\mu$, variance $\sigma^2$, the length-scale parameter vector $\nuvec$ and the nugget $g$, the Gaussian process for the data $\{\yvec, \xmat\}$ is that $\yvec$ follows a multivariate normal distribution
	$	\yvec\sim\mathbf{N}\left[\mu\onevec_n,\sigma^2\omegamat(\nuvec,g)\right],$
	where $\onevec_n$ is an $n$-element vector with all ones. The construction for the matrix $\omegamat(\nuvec)$ is the distance-inverse kernel as follows,
	\begin{align*}
		\omegamat(\nuvec,g)_{ii'} &= \exp\left[-d(\xvec_i,\xvec_{i'},\nuvec,g)\right],
\text{ and }
		d(\xvec_i,\xvec_{i'},\nuvec,g) = \sum_{l=1}^p\frac{(x_{il}-x_{i'l})^2}{\nu_l}+g\delta_{ii'}.
	\end{align*}
Here $\delta_{ii'}$ is the Kronecker delta.	The parameters $\mu$, $\sigma^2$, and $\nuvec=(\nu_1,\dots, \nu_p)^\T$ can be estimated through the maximum likelihood estimation (MLE) procedure.
	
	For prediction, the joint distribution for $\yvec$ and $\yvec_{0}$ is
	\begin{align*}
		\begin{pmatrix}
			\yvec\\
			\yvec_{0}
		\end{pmatrix}\sim\mathbf{N}
		\Bigg\{
		\begin{pmatrix}
			\mu\onevec_n\\
			\mu\onevec_{n_0}
		\end{pmatrix},
		\sigma^2
		\begin{bmatrix}
			\omegamat & \omegamat_{12}\\
			\omegamat^\T_{12} & \omegamat_{0}
		\end{bmatrix}
		\Bigg\},
	\end{align*}
	where
	$
		\left(\omegamat_{12}\right)_{ii'} = \exp\left[-d(\xvec_i,\xvec_{0i'},{\nuvec},g)\right],
		\left(\omegamat_{0}\right)_{ii'} = \exp\left[-d(\xvec_{0i},\xvec_{0i'},{\nuvec},g)\right],
	$
	$n_0$ is the number of predicted points, and $\xvec_{0i'}$ is the $i'$th input variable in the set of predicted points. The prediction for $\yvec_{0}$ is the conditional mean
	\begin{align*}
		\widehat{\yvec}_{0} = \E\left[\yvec_0|\yvec\right]= \mu\onevec_{n_0}+\omegamat_{12}^\T({\nuvec})\omegamat^{-1}({\nuvec})(\yvec-\mu\onevec_n).
	\end{align*}
	Estimation of parameters $\nuvec$ and $\mu$ will be discussed in Section~\ref{sec:estpara}.
	
	\subsection{The Linear Mixed Gaussian Process}\label{sec:lmgp}
	We construct $d$ separate models for the columns of matrix $\wmat$. For each model, we fix $j$ and use the data $\{\wvec_j,\xmat,\zmat\}$ to build the model. We consider the following model for the $\wvec_j$,
	\begin{align}\label{eqn:wij.model}
		w_{ij}&=\mu+\alpha_{ij}+\veps_{ij}, \quad i=1,\ldots,n,
	\end{align}
	where $\mu$ is the grand mean, $\alpha_{ij}$ is the categorical random effect, and $\veps_{ij}$ is the random error. For notation convenience, we drop the index $j$ but keep in mind that the model in \eqref{eqn:wij.model} will be applied separately for $j=1,\ldots, d$. With the dropping of index $j$, the model in \eqref{eqn:wij.model} is represented as
	\begin{align}\label{eqn:wij.model.without.j}
		w_{i}&=\mu+\alpha_{i}+\veps_{i}, \quad i=1,\ldots,n,
	\end{align}
	and the vector formulation of \eqref{eqn:wij.model.without.j} is
	\begin{align*}
		\wvec&=\muvec+\alphavec+\vepsvec,
	\end{align*}
	where $\wvec=\wvec_j$, $\muvec=(\mu, \ldots, \mu)^{\T}=\mu\onevec_n$, $\alphavec=(\alpha_1, \ldots, \alpha_n)^{\T}$, and $\vepsvec=(\veps_1, \ldots, \veps_n)^{\T}$.
	
	The random error term $\veps_i$ models the within-class correlation. Here, each class is one level of the categorical input combinations. Across classes, the $\veps_i$'s are independent. Specifically, we model $\vepsvec$ as a realization from a multivariate normal distribution $\mathbf{N}(\zerovec, \sigmamat_{\vepsvec})$, and the variance-covariance matrix for $\vepsvec$, denoted by $\sigmamat_{\vepsvec}$, is a block diagonal matrix. In particular,
	$$
	\sigmamat_{\vepsvec}=
	\begin{pmatrix}
		\sigmamat_{\vepsvec 1}                        &         &                        \\
		&                        \ddots  &                        \\
		&	                            & \sigmamat_{\vepsvec c} \\
	\end{pmatrix}=
	\sigma_{\vepsvec}^2\begin{pmatrix}
		\omegamat_{\vepsvec 1}                        &         &                        \\
		&                        \ddots  &                        \\
		&	                            & \omegamat_{\vepsvec c} \\
	\end{pmatrix}=\sigma_{\vepsvec}^2\omegamat_{\vepsvec}.
	$$
	Here, $\sigma_{\vepsvec}^2$ is the variance of $\veps_i$, and the $\omegamat_{\vepsvec k}$ are $n_k\times n_k$ correlation matrix. In particular, the correlation of $\veps_i$ and $\veps_{i'}$ is also the distance-inverse kernel which defined as
	\begin{align*}
		\left(\omegamat_{\epsilonvec}\right)_{ii'}=\corr(\veps_i, \veps_{i'})=
		\begin{cases}
			\exp\left[-\displaystyle\sum_{l=1}^p(x_{il}-x_{i'l})^2/\nu_l\right]+g\delta_{ii'},& \text{if } \zvec_i=\zvec_{i'}\\
			\hfill 0\hfill,              & \text{if } \zvec_i\neq\zvec_{i'}
		\end{cases}.
	\end{align*}

\begin{figure}
		\begin{center}
			\begin{tabular}{cc}
				\includegraphics[width=.45\textwidth]{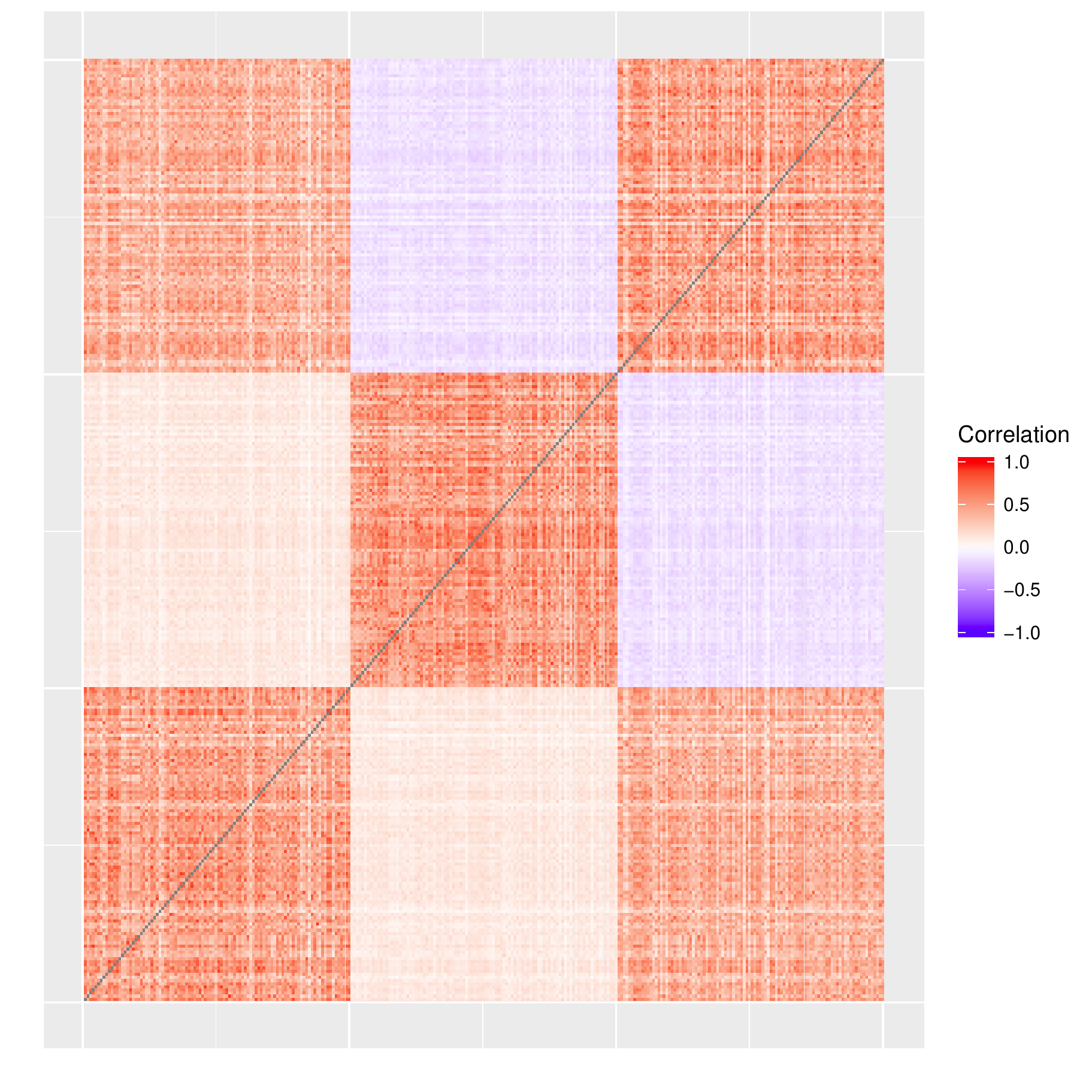}&
				\includegraphics[width=.45\textwidth]{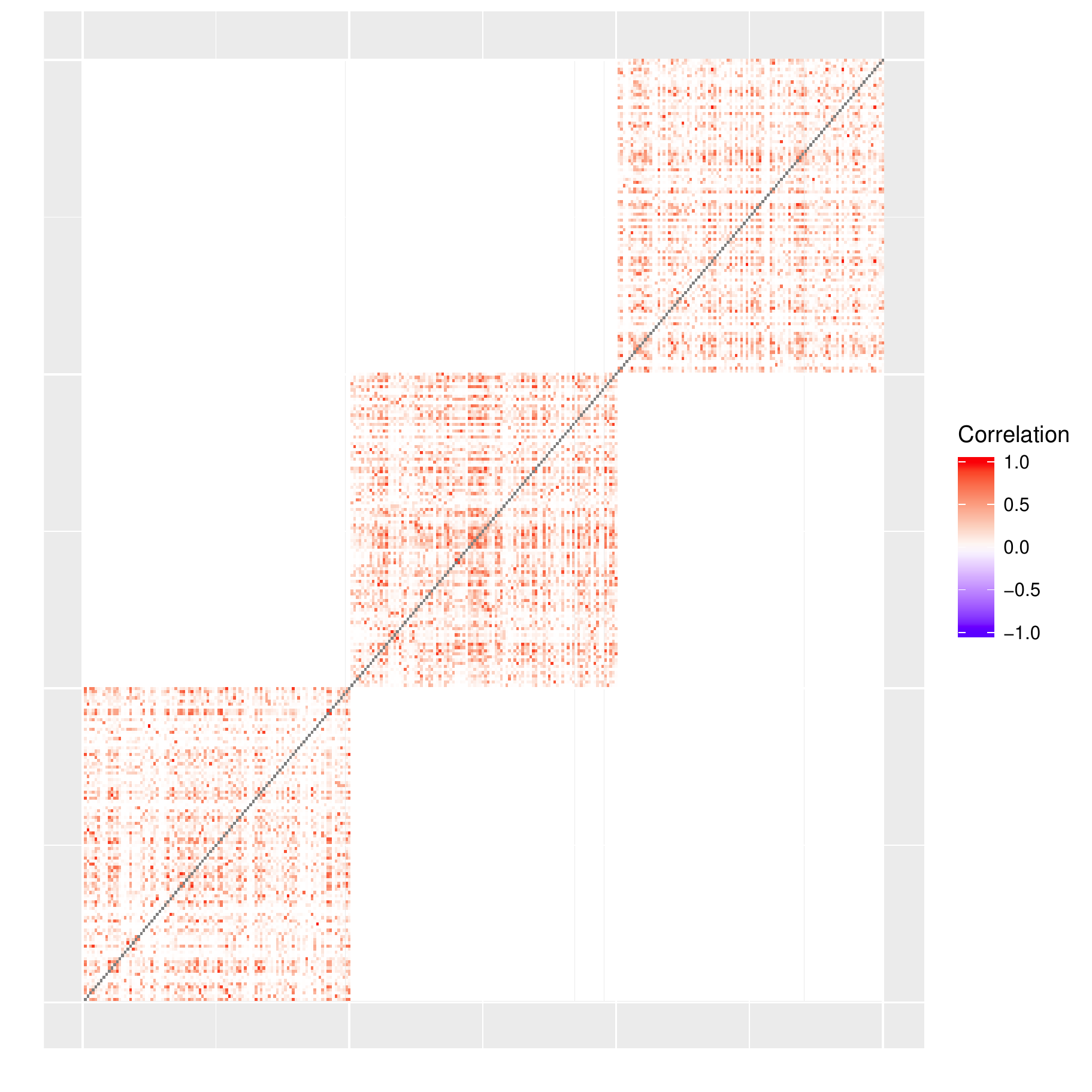}\\
				(a) $\omegamat_{\alphavec}$ & (b) $\omegamat_{\epsilonvec}$ \\
			\end{tabular}
		\end{center}
		\caption{Example correlation matrix heatmaps for $\omegamat_{\alphavec}$ and $\omegamat_{\epsilonvec}$, when $p=4$, $c=3$, $\rhovec= (1.33,0.56,2.66)^{\T}$, and $\nuvec=(2.0, 2.0, 2.0, 2.0)^{\T}$ based on the randomly generated data.}\label{fig:matheat}
	\end{figure}

	The term $\alpha_{i}$ is the categorical random effect which models the between-class correlation. We also model $\alphavec$ with a multivariate normal distribution $\mathbf{N}(\zerovec, \sigmamat_{\alphavec})$. Let $\sigmamat_{\alphavec}=\sigma_{\alphavec}^2\omegamat_{\alphavec}$, where $\sigma_{\alphavec}^2$ is the variance of $\alpha_i$, and $\omegamat_{\alphavec}$ is the corresponding correlation matrix. The structure of $\omegamat_{\alphavec}$ is specified as follows,
	\begin{align}
		\corr(\alpha_i, \alpha_{i'})=
		\begin{cases}
			\rho(\zvec_i, \zvec_{i'})\kappa(r_{ii'}, r_{\text{max}}),& \text{if } r_{ii'}\leq r_{\text{max}}\\
			\hfill 0 \hfill,              & \text{if } r_{ii'}>r_{\text{max}}
		\end{cases},\label{eq:cateconv}
	\end{align}
where $r_{ii'}=||\xvec_i-\xvec_{i'}||$ is the Euclid distance between $\xvec_i$ and $\xvec_{i'}$, $\rho(\zvec_i, \zvec_{i'})$ defines the correlation between category $i$ and $i'$, and $\kappa(\cdot, r_{\text{max}})$ is a compact support kernel with prespecified range parameter $r_{\text{max}}$ to allow for sparsity. The formula of the compact support kernel we use is
	\begin{align*}
		\kappa(r_{ii'}, r_{\text{max}})=\left(1-\frac{r_{ii'}}{r_{\text{max}}}\right)^{v}_+,\,\text{where }v\leq\frac{p+1}{2},
	\end{align*}
	which is defined by \shortciteN{wendland1995piecewise}. This functional form ensures that the $\omegamat_{\alphavec}$ is positive definite, as required for variance-covariance matrices. Let $\rho_{kk'}=\rho(\zvec_i, \zvec_{i'})$, where $k$ and $k'$ are the corresponding coded class labels for $\zvec_i$ and $\zvec_{i'}$, respectively. We use the formulation in  \shortciteN{simonian2010most} and \shortciteN{zhou2011simple} for $\rho_{kk'}$. Note that the total number of categorical level combinations is $c$. To ensure that the matrix defined by using (\ref{eq:cateconv}) is a valid variance-covariance matrix, the $c\times c$ matrix $\Pmat=(\rho_{kk'})$ must be a positive definite matrix with unit diagonal values. Let
	$
		\Pmat = \Lmat\Lmat^\T,
	$
	where $\Lmat=(l_{rs})$ is a lower triangle matrix with positive diagonal values. Let $l_{11}=1$ and for $k=2,\dots,c$ the formula for $k$th row of $\Lmat$ is given as
\[
l_{k1}=\cos(\theta_{k1}),
		l_{ks}=\left[\prod_{j=1}^{s-1}\sin(\theta_{kj})\right]\cos(\theta_{ks}),\,\text{for } s=2,\dots,k-1,
\text{ and }
		l_{kk}=\prod_{j=1}^{k-1}\sin(\theta_{kj}).
\]
	The parameters for $\omegamat_{\alpha}$ is $\theta_{ks}\in(0,\pi),\,k=2,\dots,c,\,s=1,\dots,k-1$. As a result, for $c$ categorical levels, we have $c\times (c-1)/2$ parameters for $\rho_{kk'}$. To visualize the model variance-covariance matrix structure, we provide the heatmap of a typical $\omegamat_{\alphavec}$ and $\omegamat_{\epsilonvec}$ in Figure~\ref{fig:matheat}. From Figure~\ref{fig:matheat}(b), the distance inverse kernel can only model positive correlation, which is suitable for the data within the same categorical variable combination. The $\omegamat_{\alphavec}$ in Figure~\ref{fig:matheat}(a) can model the negative correlation (the block in the top center area) between different categorical variables.

\subsection{The Estimation Procedure}\label{sec:estpara}
	Let $\thetavec_{\alphavec}=(\sigma_{\alphavec}^2, \rhovec^{\T})^{\T}$ and $\thetavec_{\vepsvec}=(\sigma_{\vepsvec}^2, \mu, \nuvec^{\T}, g)^{\T}$. All the parameters are denoted by $\thetavec=(\thetavec_{\vepsvec}^{\T}, \thetavec_{\alphavec}^{\T})^{\T}$, $\sigmamat_{\alphavec}=\sigmamat_{\alphavec}(\thetavec_{\alphavec})$, $\sigmamat_{\epsilonvec} = \sigmamat_{\epsilonvec}(\thetavec_{\vepsvec})$. The complete likelihood is
	\begin{align}
		L(\thetavec;\wvec,\alphavec)=f(\wvec,\alphavec;\thetavec)=f(\wvec|\alphavec; \thetavec)f(\alphavec;\thetavec).\label{eq:likeli}
	\end{align}
	Note that,
	\begin{align*}
		f(\wvec,\alphavec;\thetavec)&=\frac{1}{{(2\pi)^{n}}|\sigmamat_{\vepsvec}|^{1/2}|\sigmamat_{\alphavec}|^{1/2}}\exp\left\{-\frac{1}{2}
		(\wvec-\alphavec-\muvec)^{\T}\sigmamat_{\vepsvec}^{-1}(\wvec-\alphavec-\muvec)-\frac{1}{2}\alphavec^{\T}
		\sigmamat_{\alphavec}^{-1}\alphavec\right\}\\
		&\propto \exp[g(\alphavec)],
	\end{align*}
	where
$g(\alphavec)=(\wvec-\muvec)^{\T}\sigmamat_{\vepsvec}^{-1}\alphavec-\alphavec^{\T}(\sigmamat_{\vepsvec}^{-1}+
		\sigmamat_{\alphavec}^{-1})\alphavec/2.
$
Note that $\sigmamat_{\vepsvec}$ is a block diagonal matrix, and its inverse can be obtained relatively easily, and $\sigmamat_{\alphavec}$ can be a sparse matrix when data size is large. We use an EM procedure to do the estimation. The advantage is that the procedure is scalable to sample size $n$ and parameters can be estimated separately, which can reduce the difficulty of optimization.
	
	\subsubsection{Expectation Step}\label{sec:e}
	In the expectation step (E-step), at the $t$th iteration, we have $\thetavec^{(t-1)}=\left\{\left[\thetavec_{\vepsvec}^{(t-1)}\right]^\T, \left[\thetavec_{\alphavec}^{(t-1)}\right]^\T\right\}^{\T}$. Let $\sigmamat_{\alphavec}\left[\thetavec_{\alphavec}^{(t-1)}\right] = \sigmamat_{\alphavec}^{(t-1)}$ and $\sigmamat_{\epsilonvec}\left[\thetavec_{\epsilonvec}^{(t-1)}\right] = \sigmamat_{\epsilonvec}^{(t-1)}$. The expectation is
	\begin{align*}
		\mathcal{Q}\left[\thetavec|\thetavec^{(t-1)}\right]=\E_{\alphavec|\wvec,\thetavec_{\vepsvec}^{(t-1)},\thetavec_{\alphavec}^{(t-1)}}\log\left[L(\thetavec;\wvec,\alphavec)\right].
	\end{align*}
	We need to derive the distribution of $\alphavec|\wvec,\thetavec_{\vepsvec}^{(t-1)},\thetavec_{\alphavec}^{(t-1)}$. The joint distribution for $\wvec$ and $\alphavec$ given $\thetavec_{\vepsvec}^{(t-1)}$ and $\thetavec_{\alphavec}^{(t-1)}$ is
	\begin{align*}
		\begin{bmatrix}
			\alphavec|\thetavec_{\alphavec}^{(t-1)}\\
			\wvec|\thetavec_{\epsilonvec}^{(t-1)}
		\end{bmatrix}
		&\sim \mathbf{N}
		\Bigg\{
		\begin{bmatrix}
			\mathbf{0}_n\\
			\muvec^{(t-1)}
		\end{bmatrix},
		\begin{bmatrix}
			\sigmamat_{\alphavec}^{(t-1)} & \sigmamat_{\alphavec}^{(t-1)}\\
			\sigmamat_{\alphavec}^{(t-1)} & \sigmamat_{\alphavec}^{(t-1)}+\sigmamat_{\vepsvec}^{(t-1)}
		\end{bmatrix}
		\Bigg\},
	\end{align*}
	where $\mathbf{0}_n$ is an $n$-element vector with all zero entries.
	By the properties of normal distribution, the distribution of $\alphavec|\wvec,\thetavec_{\vepsvec}^{(t-1)},\thetavec_{\alphavec}^{(t-1)}$ is normal with the mean and covariance matrix as
	\begin{align*}
		\E\left[\alphavec|\wvec,\thetavec_{\vepsvec}^{(t-1)},\thetavec_{\alphavec}^{(t-1)}\right] =
		\sigmamat_{\alphavec}^{(t-1)}\left[\sigmamat_{\alphavec}^{(t-1)}+\sigmamat_{\vepsvec}^{(t-1)}\right]^{-1}\left[\wvec-\muvec^{(t-1)}\right],\\
		\text{Cov}\left[\alphavec|\wvec,\thetavec_{\vepsvec}^{(t-1)},\thetavec_{\alphavec}^{(t-1)}\right] = \sigmamat_{\alphavec}^{(t-1)}-\sigmamat_{\alphavec}^{(t-1)}
\left[\sigmamat_{\alphavec}^{(t-1)}+\sigmamat_{\vepsvec}^{(t-1)}\right]^{-1}\sigmamat_{\alphavec}^{(t-1)},
	\end{align*}
respectively. Here, to ensure model estimability, we introduce the zero-sum constraint for $\alphavec$, that is $\sum_{i=1}^n\alpha_i=0$. To achieve this, we multiply $\alphavec$ by a centering matrix $\cmat=\mathbf{I}-n^{-1}\mathbf{J}_n$, where $\mathbf{J}_n$ is an $n\times n$ all-ones matrix. The centralized $\alphavec$ has a singular multivariate normal distribution with mean and covariance
	\begin{align*}
		\E\left[\alphavec|\wvec,\thetavec_{\vepsvec}^{(t-1)},\thetavec_{\alphavec}^{(t-1)}\right] &= \cmat
		\sigmamat_{\alphavec}^{(t-1)}\left[\sigmamat_{\alphavec}^{(t-1)}+\sigmamat_{\vepsvec}^{(t-1)}\right]^{-1}\left[\wvec-\muvec^{(t-1)}\right],\\
		\text{Cov}\left[\alphavec|\wvec,\thetavec_{\vepsvec}^{(t-1)},\thetavec_{\alphavec}^{(t-1)}\right] &= \cmat\left\{\sigmamat_{\alphavec}^{(t-1)}-\sigmamat_{\alphavec}^{(t-1)}
\left[\sigmamat_{\alphavec}^{(t-1)}+\sigmamat_{\vepsvec}^{(t-1)}\right]^{-1}
\sigmamat_{\alphavec}^{(t-1)}\right\}\cmat,
	\end{align*}
respectively. Expanding the complete likelihood function in (\ref{eq:likeli}), we obtain
	\begin{align*}
		\loglik(\thetavec)=\log[L(\thetavec;\wvec,\alphavec)]=\loglik_{1}(\thetavec_{\vepsvec}; \wvec|\alphavec)+\loglik_{2}(\thetavec_{\alphavec};\alphavec).
	\end{align*}
	Here,
	\begin{align*}
		\loglik_{1}(\thetavec_{\vepsvec}; \wvec|\alphavec)=-\frac{n}{2}\log(2\pi)-\frac{1}{2}\log(|\sigmamat_{\vepsvec}|)-\frac{1}{2}
		(\wvec-\muvec-\alphavec)^{\T}\sigmamat_{\vepsvec}^{-1}(\wvec-\muvec-\alphavec),
	\end{align*}
	and
	\begin{align*}
		\loglik_{2}(\thetavec_{\alphavec};\alphavec)=-\frac{n}{2}\log(2\pi)-\frac{1}{2}\log(|\sigmamat_{\alphavec}|)-\frac{1}{2}
		\alphavec^{\T}\sigmamat_{\alphavec}^{-1}\alphavec.
	\end{align*}
	Taking the expectation with respect to $\alphavec$, we have
	\begin{align}
		\mathcal{Q}_1\left[\thetavec_{\epsilonvec}|\thetavec_{\epsilonvec}^{(t-1)}\right]=&
		-\frac{n}{2}\log(2\pi)-\frac{1}{2}\log(|\sigmamat_{\epsilonvec}|)-\frac{1}{2}(\wvec-\muvec)^{\T}\sigmamat_{\vepsvec}^{-1}(\wvec-\muvec)\nonumber\\
		&+(\wvec-\muvec)^{\T}\sigmamat_{\epsilonvec}^{-1}\muvec_{\alphavec|\wvec}^{(t-1)}-\frac{1}{2}\tr\left[\sigmamat_{\epsilonvec}^{-1}\sigmamat_{\alphavec|\wvec}^{(t-1)}\right]-\frac{1}{2}\left[\muvec_{\alphavec|\wvec}^{(t-1)}\right]^{\T}\sigmamat_{\epsilonvec}^{-1}\muvec_{\alphavec|\wvec}^{(t-1)}\label{eq:q1},
	\end{align}
	and
	\begin{align*}
		\mathcal{Q}_2\left[\thetavec_{\alphavec}|\thetavec_{\alphavec}^{(t-1)}\right]=-\frac{n}{2}\log(2\pi)-\frac{1}{2}\log(|\sigmamat_{\alphavec}|)-\frac{1}{2}\tr\left[\sigmamat_{\alphavec}^{-1}\sigmamat_{\alphavec|\wvec}^{(t-1)}\right]-\frac{1}{2}\left[\muvec_{\alphavec|\wvec}^{(t-1)}\right]^{\T}\sigmamat_{\alphavec}^{-1}\muvec_{\alphavec|\wvec}^{(t-1)}.
	\end{align*}
	The derivation for $\mathcal{Q}_1\left[\thetavec_{\epsilonvec}|\thetavec_{\epsilonvec}^{(t-1)}\right]$ and $\mathcal{Q}_2\left[\thetavec_{\alphavec}|\thetavec_{\alphavec}^{(t-1)}\right]$ are provided in Appendix~\ref{sec:ch4ap1}.
	
	\subsubsection{M-Step for Parameter Estimation}\label{sec:m}
	The updating formulas for $\thetavec_{\epsilonvec}$ and $\thetavec_{\alphavec}$ are
	\begin{align*}
		\thetavec_{\epsilonvec}^{(t)} &= \arg\max_{\thetavec_{\epsilonvec}}\mathcal{Q}_1
\left[\thetavec_{\epsilonvec}|\thetavec_{\epsilonvec}^{(t-1)}\right]
\text{ and }
		\thetavec_{\alphavec}^{(t)} = \arg\max_{\thetavec_{\alphavec}}\mathcal{Q}_2\left[\thetavec_{\alphavec}|\thetavec_{\alphavec}^{(t-1)}\right],
	\end{align*}
respectively. For $\mu$, $\sigma_{\epsilonvec}^2$, and $\sigma_{\alphavec}^2$, we have closed forms for updating as follows,
	\begin{align*} \wh\mu=&\frac{\onevec^{\T}\sigmamat_{\epsilonvec}^{-1}(\wvec-\muvec_{\alphavec|\wvec})}{\onevec^{\T}\sigmamat_{\epsilonvec}^{-1}\onevec},\\
		n\wh\sigma_{\epsilonvec}^2=&(\wvec-\wh\muvec)^{\T}\omegamat_{\vepsvec}^{-1}(\wvec-\wh\muvec)-
		2(\wvec-\wh\muvec)^{\T}\omegamat_{\epsilonvec}^{-1}\muvec_{\alphavec|\wvec}\\
		&+\tr\left[\omegamat_{\epsilonvec}^{-1}\sigmamat_{\alphavec|\wvec}^{(t-1)}\right]+\left[\muvec_{\alphavec|\wvec}^{(t-1)}\right]^{\T}\omegamat_{\epsilonvec}^{-1}\muvec_{\alphavec|\wvec}^{(t-1)},\\
		n\wh\sigma_{\alphavec}^2=&\tr\left[\omegamat_{\alphavec}^{-1}\sigmamat_{\alphavec|\wvec}^{(t-1)}\right]+\left[\muvec_{\alphavec|\wvec}^{(t-1)}\right]^{\T}\omegamat_{\alphavec}^{-1}\muvec_{\alphavec|\wvec}^{(t-1)}.
	\end{align*}
	Substituting $\wh\mu$, $\wh\sigma_{\epsilonvec}^2$, and $\wh\sigma_{\alphavec}^2$ into $\mathcal{Q}_1$ and $\mathcal{Q}_2$, we have the profile likelihood for $\nuvec$, $g$ and $\rhovec$ as follows,
	\begin{align*}
		\mathcal{Q}_1&\left[\nuvec,g|\thetavec_{\epsilonvec}^{(t-1)},\wh\mu,\wh\sigma_{\epsilonvec}^2\right]=-\frac{1}{2}\log(|\omegamat_{\epsilonvec}|)-\frac{n}{2}\log(\wh\sigma_{\epsilonvec}^2),\\
		\mathcal{Q}_2&\left[\rhovec|\thetavec_{\alphavec}^{(t-1)},\wh\mu,\wh\sigma_{\alphavec}^2\right]=-\frac{1}{2}\log(|\omegamat_{\alphavec}|)-\frac{n}{2}\log(\wh\sigma_{\alphavec}^2).
	\end{align*}
	We use the ``L-BFGS-B" in the R routine ``optim", which is a gradient-based method (\shortciteNP{zhu1995limited}), to solve the optimization problem for $\nuvec$, $g$ and $\rhovec$. We use the estimated value of $\nuvec$, $g$ and $\rhovec$ by GP as the optimization starting values.
	
	\subsubsection{Different $\mu$, $\sigma_{\epsilonvec}^2$, $g$, and $\nuvec$ for Each Category}
	The model we construct so far shares the same $\mu$, $\sigma_{\epsilonvec}^2$, $g$, and $\nuvec$ in all categories. But in some applications, it is possible that data in different categories behave differently. For example, the throughput for the I/O modes random\_reader and reader are different because random\_reader tests the speed of reading large amounts of small files while reader tests the speed of reading large files. As a result, we also provide the formula for the model with different $\mu$, $\sigma_{\epsilonvec}^2$, $g$, and $\nuvec$ separately for each category in this section. We refer to this model as LMGP-S. Note that the LMGP-S model still has correlations among different categories.
	
	For category $k$, $k=1,\dots,c$, let $\mathbb{I}_k$ be the set of indexes that all the data points belong to category $k$. In other words, $\mathbb{I}_k$ is an index set with $n_k$ elements. Let $\mu_k$, $\sigma_{\epsilonvec,k}^2$, $g_k$, and $\nuvec_k$ be the parameters for the $\Sigmavec_{\epsilonvec,k}$ in category $k$. Then updating formulas for $\mu_k$ and $\sigma_{\epsilonvec,k}^2$ in the M step are:
	\begin{align*}
		\wh\mu_k=&\frac{\onevec^{\T}_{\mathbb{I}_k}\sigmamat_{\epsilonvec,\mathbb{I}_k}^{-1}(\wvec_{\mathbb{I}_k}-\muvec_{\alphavec|\wvec,\mathbb{I}_k})}{\onevec^{\T}_{\mathbb{I}_k}\sigmamat_{\epsilonvec,\mathbb{I}_k}^{-1}\onevec_{\mathbb{I}_k}},\\
		n\wh\sigma_{\epsilonvec,k}^2=&(\wvec_{\mathbb{I}_k}-\wh\muvec_{k})^{\T}\omegamat_{\vepsvec,\mathbb{I}_k}^{-1}(\wvec_{\mathbb{I}_k}-\wh\muvec_k)-
		2(\wvec_{\mathbb{I}_k}-\wh\muvec_k)^{\T}\omegamat_{\epsilonvec,\mathbb{I}_k}^{-1}\muvec_{\alphavec|\wvec,\mathbb{I}_k}\\
		&+\tr\left[\omegamat_{\epsilonvec,\mathbb{I}_k}^{-1}\sigmamat_{\alphavec|\wvec,\mathbb{I}_k}^{(t-1)}\right]+\left[\muvec_{\alphavec|\wvec,\mathbb{I}_k}^{(t-1)}\right]^{\T}\omegamat_{\epsilonvec,\mathbb{I}_k}^{-1}\muvec_{\alphavec|\wvec,\mathbb{I}_k}^{(t-1)},
	\end{align*}
	where $\sigmamat_{\epsilonvec,\mathbb{I}_k}$, $\omegamat_{\epsilonvec,\mathbb{I}_k}$, and $\sigmamat_{\alphavec|\wvec,\mathbb{I}_k}$ are the corresponding $n_k\times n_k$ block matrix for all the data points in category $k$ in $\sigmamat_{\epsilonvec}$, $\omegamat_{\epsilonvec}$, and $\sigmamat_{\alphavec|\wvec}$. Other formulas for the EM algorithm are the same as derived before. The derivations for $\wh\mu_k$ and $\wh\sigma_{\epsilonvec,k}^2$ are provided in Appendix~\ref{sec:ch4ap2}. Some further technical details for derivatives are given in Appendix~\ref{sec:formula.score.function}.

	\subsection{Prediction for Distributional Outcomes}
	For a new configuration $(\xvec_0^{\T}, \zvec_0^{\T})^{\T}$, the goal is to predict its distribution function and we can do this by predicting $\wvec_0=(w_{01}, \ldots, w_{0d})^{\T}$. The prediction is based on those $d$ separate models in \eqref{eqn:wij.model}. Here, we describe how to make the prediction for the $j$th element of $\wvec_0$ based on the following model,
	\begin{align*}
		w_{0j}&=\mu+\alpha_{0j}+\veps_{0j}, j=1, \ldots, d.
	\end{align*}
	For notation convenience, we drop the index $j$ and work on the following model.
	\begin{align*}
		w_{0}&=\mu+\alpha_{0}+\veps_{0}.
	\end{align*}
	We construct
	\begin{align*}
		\begin{pmatrix}
			w_0\\
			\wvec
		\end{pmatrix}\sim\mathbf{N}\left[
		\begin{pmatrix}
			\mu\\
			\mu\onevec_n
		\end{pmatrix},
		\begin{pmatrix}
			\sigmamat_{00} & \sigmamat_{01}\\
			\sigmamat_{10} & \sigmamat_{11}
		\end{pmatrix}
		\right].
	\end{align*}
	With estimated $\wh\thetavec$, the predicted $\wh w_0$ is the conditional mean
	\begin{align}
		\wh w_0=\E(w_0|\wvec)=\mu+\sigmamat_{01}\left(\sigmamat_{11}\right)^{-1}(\wvec-\mu\onevec_n).\label{eqn:w.pred}
	\end{align}
	Repeat the prediction in \eqref{eqn:w.pred} for $j=1, \ldots, d'$ to obtain the prediction for $w_0$ to obtain $\wh w_0$. Here, $d'\leq d$, $d'$ is the number of SVD components and is chosen by computing budget and prediction accuracy. Another way to determine $d'$ is to let $d'$ be the smallest integer such that
	$
		{\sum_{j=1}^{d'}\lambda_j}/{\sum_{j=1}^{d}\lambda_j}\geq \text{threshold,}
	$
	which is selected to ensure sufficient modeling fidelity for predictive purposes. Let $\betavec_0=(\beta_{00}, \beta_{01},\ldots,\beta_{0d})^{\T}$. The predicted coefficients for the splines are recovered by
	\begin{align}
		\wh \betavec_0=\wh\wvec_0\vmat_{d'}^{\T},\label{eq:wpre}
	\end{align}
	according to the SVD, where $\vmat_{d'}$ is the first $d'$ columns of $\vmat$. Because it is possible that some of the elements in $\wh \betavec_0$ are negative. These negative entries are truncated to 0. The prediction for the quantile function $Q_0(p)$ is then obtained as
	\begin{align}
		\wh Q_0(p)=\wh\beta_{00}+\sum_{j=1}^{d-1}\wh\beta_{0j}\gamma_j(p).\label{eq:curpre}
	\end{align}
	Note that truncation at zero is justified because it results in the nearest point to $\wh \betavec_0$ in the convex hull that makes $\wh Q_0(p)$ a monotone function. The prediction of the CDF, $\wh F_0(y)$, can be obtained by inverting $\wh Q_0(p)$.
	
	\section{Prediction Performance Study Using HPC Data}\label{sec:res}
	We first introduce the prediction model variants and the error metric for comparisons. We then demonstrate that the SVD can reduce the dimension of the $\Bmat$ without much loss of accuracy. We compare the prediction performance of the four model variants under the proposed prediction framework. We also visualize the results of predicting quantile functions.

	\subsection{Prediction Models and Performance Metrics}
		Our prediction framework can have four variants, depending on the GP model used for predicting $\wvec$. In particular, they are

	\begin{inparaitem}
		\item LMGP: Linear mixed Gaussian process with common $\mu$, $\sigma_{\epsilonvec}^2$, $g$, and $\nuvec$ for all categories.

		\item LMGP-S: Linear mixed Gaussian process with different $\mu$, $\sigma_{\epsilonvec}^2$, $g$, and $\nuvec$ for each category.
		
\item GP: Separate simple Gaussian process fitting for each category.

		\item CGP: Categorical Gaussian process in \shortciteN{zhou2011simple}. The CGP is a modified GP using the $\sigmamat_{\alphavec}$ with no threshold for $r_{ii'}$ as the variance-covariance matrix.

\end{inparaitem}
The GP and CGP can be treated as two special cases of the LMGP. Based on the four model variants, the quantile function $Q_0(p)$ can be predicted using (\ref{eq:wpre}) and (\ref{eq:curpre}). Because there is no existing methods for comparisons, we compare the prediction accuracy under the four model variants.

The prediction accuracy is measured by comparing the discrepancy between the smoothed CDF and predicted CDF. Let $F(y)$ be the sample CDF and $\widehat{F}(y)$ be the predicted CDF; we use the errors based on the $L^1$-norm $(EL_1)$ for error measurement:
	\begin{align*}
		EL_1 = ||F(y)-\widehat{F}(y)||_1.
	\end{align*}
	All the $EL_1$'s in this paper are on the scale of $10^7$ KB/s.

	We show the prediction accuracy for different prediction tasks on IOzone data. The algorithms are implemented in R (\shortciteNP{R}). Because our focus is on prediction, we test the prediction framework on real datasets, instead of using simulated datasets. To create multiple datasets for training/testing purposes, we obtain subsets with three I/O modes from the IOzone database.

	\subsection{Dimension Reduction by Selection of $d'$}
\begin{figure}
		\begin{center}
			\begin{tabular}{cc}
				\includegraphics[width=.47\textwidth,angle=270,origin=c]{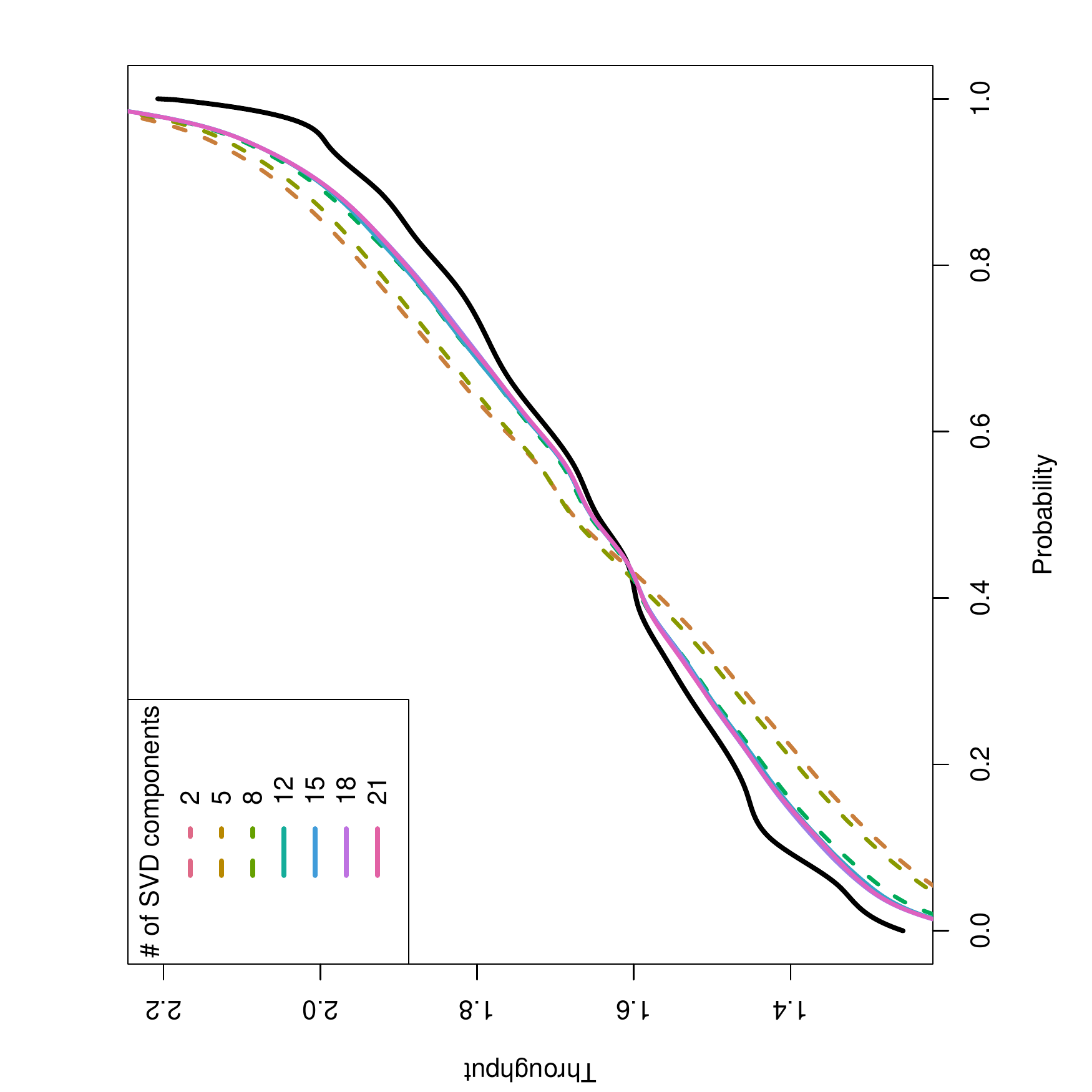}&
				\includegraphics[width=.47\textwidth,angle=270,origin=c]{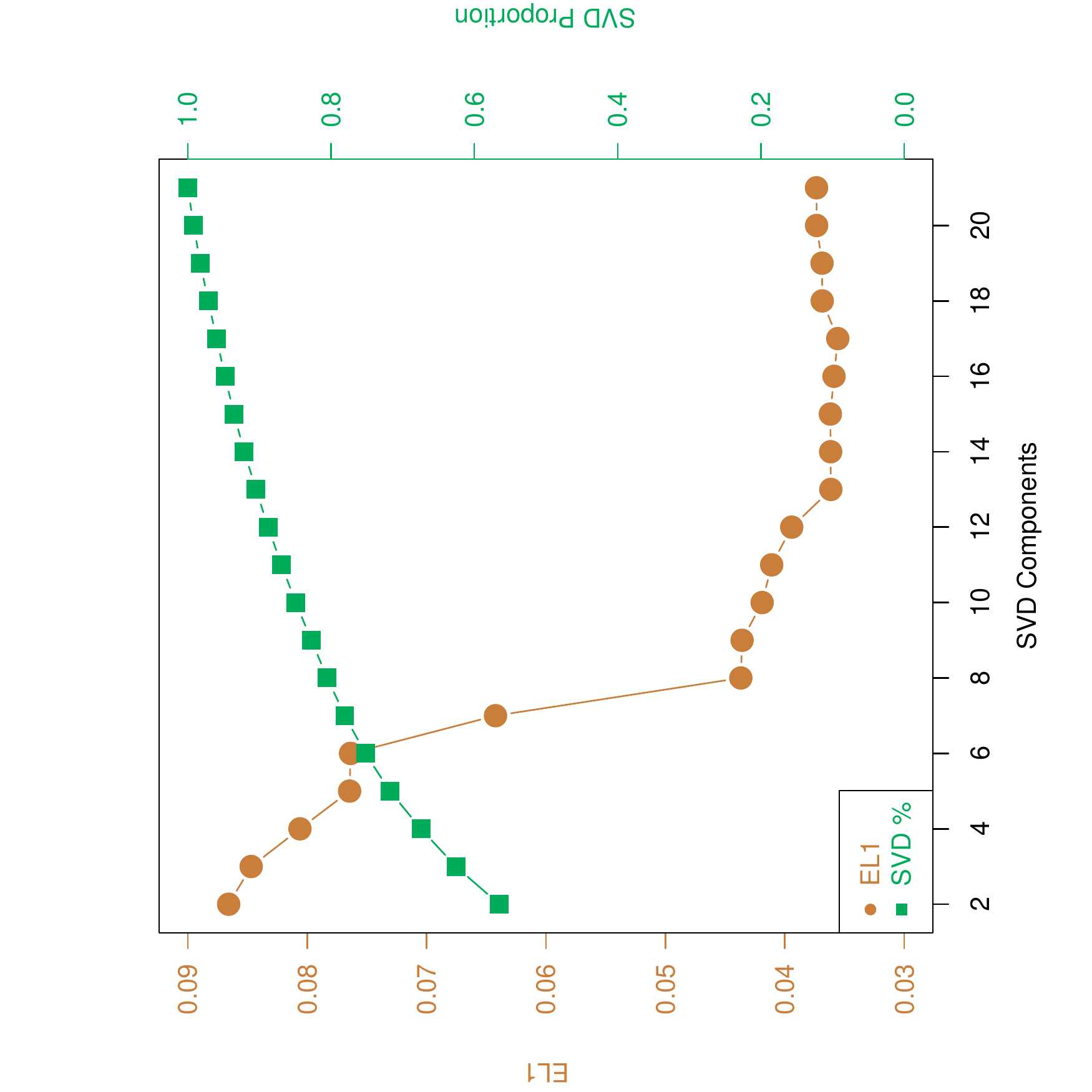}\\
				(a) \multilinecell{Predicted Quantile}  & (b) \multilinecell{$EL_1$ and SVD Proportion}\\
			\end{tabular}
		\end{center}
		\caption{Prediction under different numbers of the SVD components. The black solid line in Figure (a) is the smoothed sample quantile function.}\label{fig:svdcomp}
	\end{figure}

	In this section, we show the model's potential in reducing the dimension by selecting the number of SVD components $d'$. The dataset used here contains three modes: random\_writer, rereader, and reader. We randomly choose 20\% of the data as the test set. Figure~\ref{fig:svdcomp} shows the predicted curves by LMGP using different numbers of SVD components. In Figure~\ref{fig:svdcomp}(a), when we use more than 12 components, the predicted curves (solid lines) are quite similar and are very close to the true black quantile function. This observation is also confirmed by Figure~\ref{fig:svdcomp}(b). The decreasing trend of the $EL_1$ vanishes when the number of components increases beyond eight, where about 80\% of the singular values are covered.
	Thus the dimension of $\betavec$ can be reduced by SVD without much loss of accuracy. In the rest of this paper, all of our predictions are based on the first 12 SVD components (i.e., $d'=12$).

	\subsection{Average Error for Different Training Set Proportions}
	In this section, we discuss the prediction accuracy on different training set proportions. We create five datasets and each dataset is a different combination of the three IO operation modes (the categorical input) from the large IOzone database. For each dataset, the training proportions are from 30\% to 70\%. To obtain the average error, the random train-test splitting is repeated 100 times. The results for five datasets are shown in Table~\ref{tab:el1main}.
	
	\begin{figure}
		\begin{center}
			\begin{tabular}{cc}
				\includegraphics[width=.45\textwidth,angle=270,origin=c]{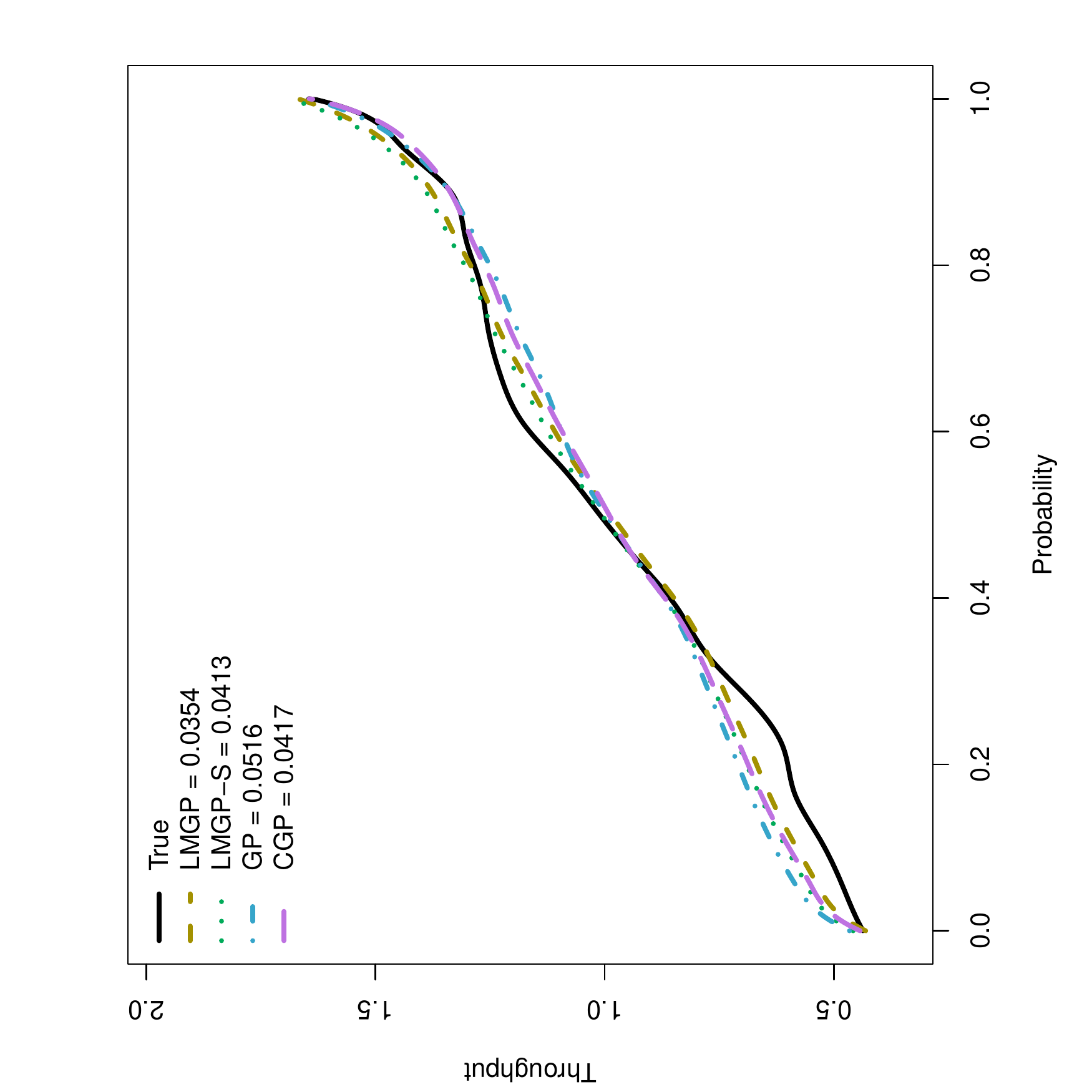} &
				\includegraphics[width=.45\textwidth,angle=270,origin=c]{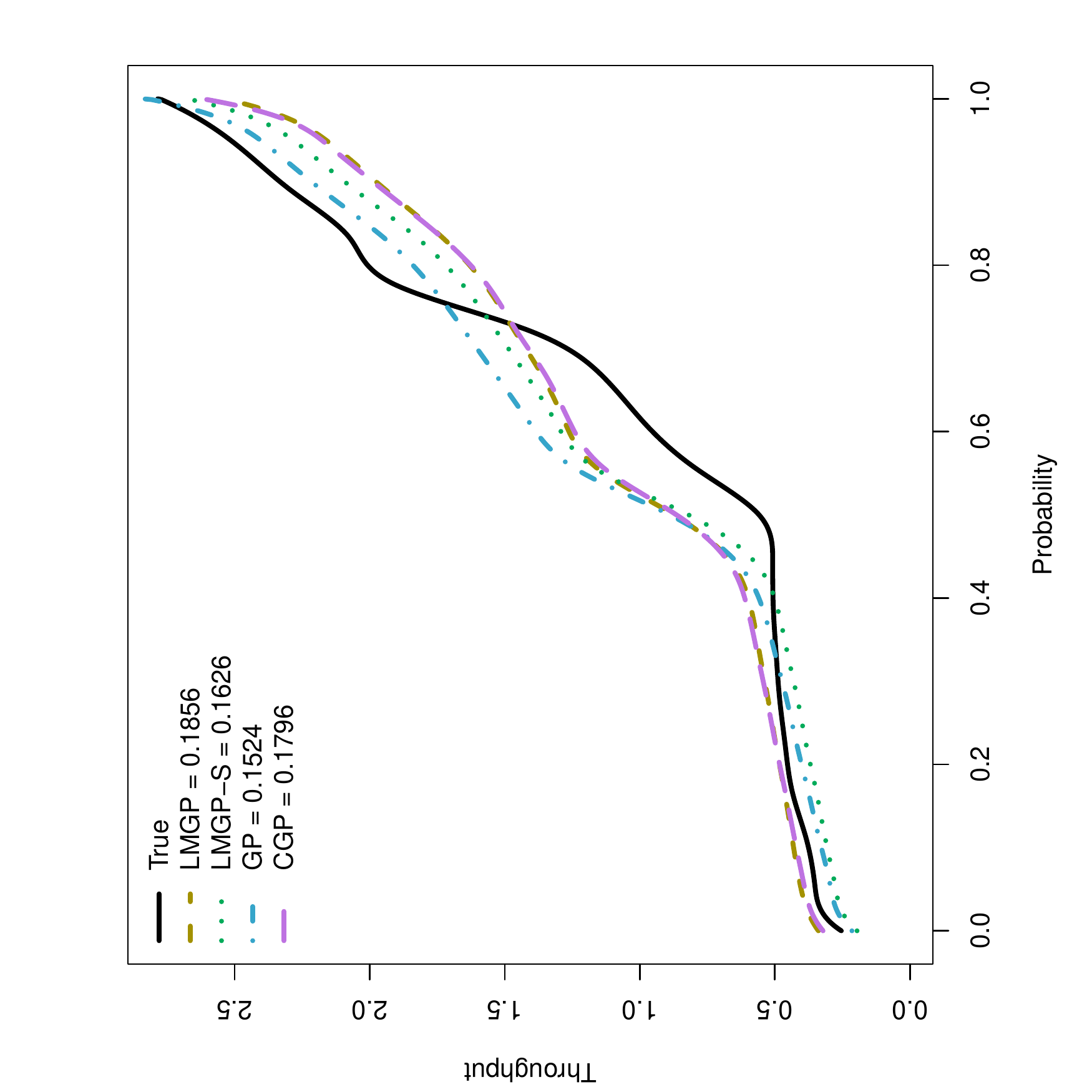} \\
				(b) \multilinecell{Prediction for an Interior Point} &
				(a) \multilinecell{Prediction for a Boundary Point}
			\end{tabular}
		\end{center}
		\caption{Two examples of the predicted quantile functions by using the four models. The $\xvec$ for the left panel is an interior point (Freq=2.3, FS=8192, Thread=24) in the training set while the $\xvec$ for the right panel is at the boundary (Freq=3.5, FS=1024, Thread=8). The legends show the $EL_1$ for each model.}\label{fig:qtpred}
	\end{figure}
	
From Table~\ref{tab:el1main}, we can see that the prediction accuracy generally increases when the proportion of the dataset used for training increases. For most cases, the LMGP-S model variant has the best performance, the LMGP variant is the next one, and the performance of GP and CGP is worse than the LMGP-S, which reveals that there are correlations among data in different I/O modes. In some cases, the GP model variant has the best performance and the performance of LMGP-S is close to that of GP. Overall, the LMGP-S model variant provides the most consistently accurate results.

To visualize the prediction results, we provide two examples of the predicted quantile functions in the test set as shown in Figure~\ref{fig:qtpred}. When the test point is an interior point of the training set (e.g., as the point shown in Figure~\ref{fig:qtpred}(a)), the predicted quantile functions are quite close to the sample quantile. The predicted curves of LMGP and LMGP-S are closer to the true curve compared with CGP and GP. When the test point is close to the boundary (as shown in Figure~\ref{fig:qtpred}(b)), the prediction is poor. GP-based models are intended for interpolation (i.e., the $\xvec_0$ is inside the convex hull of the data). When $\xvec_0$ is near the boundary or outside the convex hull of the data, the performance tends to be poor.

	\begin{table}
		\centering
		\caption{Average $EL_1$ after multiple train-test splits on multiple datasets with different 3-mode combinations.}\label{tab:el1main}
\begin{center}
		\begin{tabular}{c|crrrr}
			\hline
			\multirow{2}{*}{Modes in Dataset}&\multirow{2}{*}{\multilinecell{Training\\Proportion}}& \multicolumn{4}{c}{$EL_1$} \\\cline{3-6}
			&& LMGP & LMGP-S & GP & CGP   \\\hline
			\multirow{5}{*}{\multilinecell{random\_reader\\random\_writer\\rereader}}&0.3 & 0.0428 & 0.0427 & 0.0472 & 0.0449 \\
			&0.4 & 0.0387 & 0.0385 & 0.0430 & 0.0405 \\
			&0.5 & 0.0367 & 0.0363 & 0.0411 & 0.0381 \\
			&0.6 & 0.0347 & 0.0342 & 0.0389 & 0.0360 \\
			&0.7 & 0.0340 & 0.0337 & 0.0382 & 0.0353 \\
			\hline
			\multirow{5}{*}{\multilinecell{random\_writer\\rereader\\reader}}&0.3 & 0.0423 & 0.0421 & 0.0465 & 0.0443 \\
			&0.4 & 0.0381 & 0.0380 & 0.0428 & 0.0400 \\
			&0.5 & 0.0355 & 0.0349 & 0.0404 & 0.0371 \\
			&0.6 & 0.0339 & 0.0335 & 0.0388 & 0.0359 \\
			&0.7 & 0.0347 & 0.0340 & 0.0375 & 0.0349 \\
			\hline
			\multirow{5}{*}{\multilinecell{rereader\\reader\\rewriter}}&0.3 & 0.0422 & 0.0417 & 0.0453 & 0.0442 \\
			&0.4 & 0.0381 & 0.0377 & 0.0421 & 0.0399 \\
			&0.5 & 0.0360 & 0.0355 & 0.0401 & 0.0375 \\
			&0.6 & 0.0345 & 0.0337 & 0.0384 & 0.0356 \\
			&0.7 & 0.0343 & 0.0330 & 0.0370 & 0.0342 \\
			\hline
			\multirow{5}{*}{\multilinecell{initial\_writer\\random\_reader\\random\_writer}}&0.3 & 0.0319 & 0.0305 & 0.0307 & 0.0340 \\
			&0.4 & 0.0298 & 0.0292 & 0.0278 & 0.0305 \\
			&0.5 & 0.0294 & 0.0268 & 0.0258 & 0.0282 \\
			&0.6 & 0.0288 & 0.0252 & 0.0248 & 0.0268 \\
			&0.7 & 0.0297 & 0.0257 & 0.0239 & 0.0258 \\
			\hline
			\multirow{5}{*}{\multilinecell{initial\_writer\\random\_writer\\rereader}}&0.3 & 0.0285 & 0.0295 & 0.0285 & 0.0310 \\
			&0.4 & 0.0266 & 0.0307 & 0.0267 & 0.0288 \\
			&0.5 & 0.0256 & 0.0283 & 0.0252 & 0.0269 \\
			&0.6 & 0.0245 & 0.0257 & 0.0239 & 0.0252 \\
			&0.7 & 0.0255 & 0.0261 & 0.0235 & 0.0248 \\
			\hline
		\end{tabular}
\end{center}
	\end{table}
\section{Predicting Summary Statistics and Comparisons}\label{sec:sumstat}
One application of distributional predictions is to predict the summary statistics of a distribution from the predicted quantile function/CDF. Typical summary statistics can be the mean, standard deviation (SD), and quantile values of the underlying distribution. For predicting summary statistics, there are also existing methods available. Thus, we make comparisons with existing methods in predicting summary statistics in this section.

\begin{figure}
		\begin{center}
			\includegraphics[width=.85\textwidth]{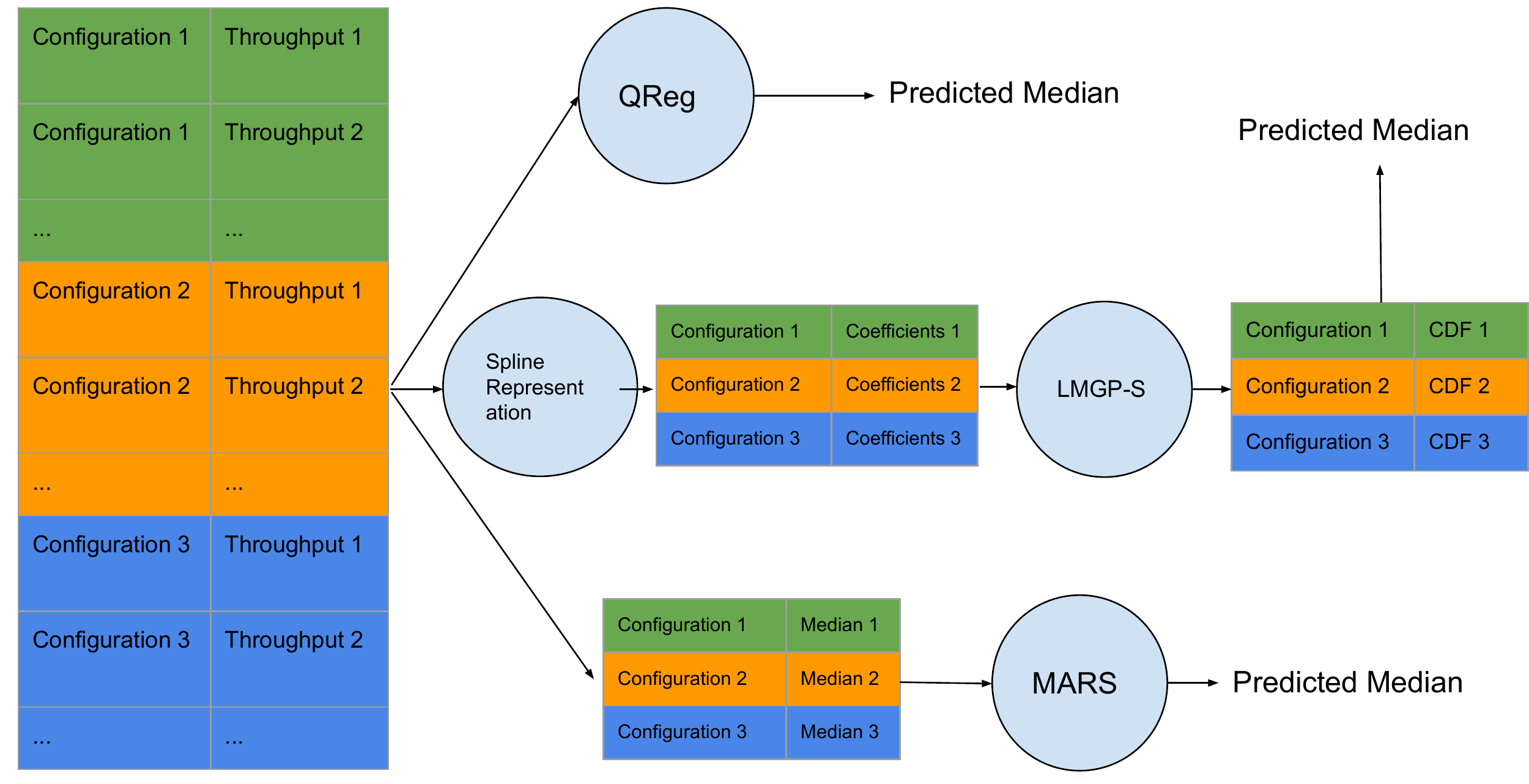}
		\end{center}
		\caption{Flow chart that illustrates how data are processed before being fed to different models for median prediction.}\label{fig:workflow}
	\end{figure}

\shortciteN{xu2020prediction} study the accuracy of predicting throughput standard deviations using multiple surrogates. For comparison, we have two baseline models which can incorporate both quantitative and qualitative factors. The first baseline model is quantile regression~(\shortciteNP{10.2307/1913643,li2021tensor}). Quantile regression (QReg) can predict the quantile of the throughput given all the replicated throughputs. The other comparison method is MARS. We use the R package ``earth" (\shortciteNP{Rearth}) for implementing MARS and ``quantreg" (\shortciteNP{RQReg}) for implementing QReg. The summary statistics are calculated from the throughputs under a given configuration. Figure~\ref{fig:workflow} provides a flow chart on how data are processed before being fitted to different models. The QReg can take the raw data with replicated throughputs and predict the median directly. For MARS, the sample median is calculated and then used in model training. We use the LMGP-S model variant here. For LMGP-S, the median is calculated from the predicted quantile function. The summary statistic of interest in Figure~\ref{fig:workflow} is the median (i.e., the 0.5 quantile).

\begin{figure}
		\begin{center}
			\includegraphics[width=0.75\textwidth,angle=270,origin=c]{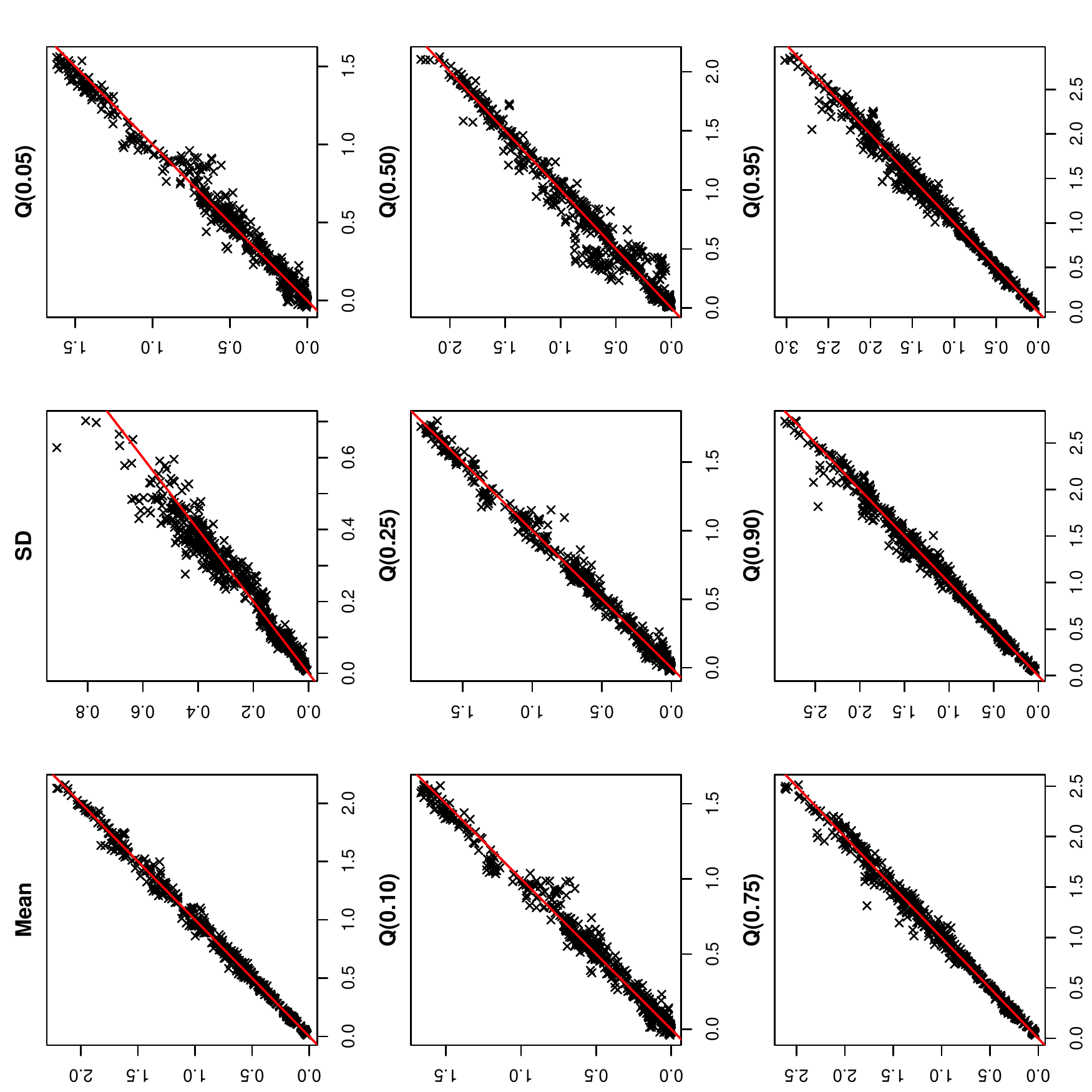}
		\end{center}
		\caption{Scatter plots of several predicted summary statistics using LMGP-S. The $x$-axis is the predicted value and the $y$-axis is the true value.}\label{fig:SumStatlmgp}
	\end{figure}

	The dataset we used for summary statistics prediction has three I/O modes: random\_writer, rereader, and reader. 20\% of the data are randomly chosen to be the test set. The error measure for the summary statistics (a scalar output) is the mean squared error (MSE).  Figure~\ref{fig:SumStatlmgp} shows the scatter plots between the LMGP-S predicted and true summary statistics on the test set. Figure~\ref{fig:SumStatlmgp} shows the LMGP-S has accurate predictions even for the 0.05 and 0.95 quantiles. Almost all the points are close to the $y=x$ line. Table~\ref{tab:SumStat} shows the MSEs for different summary statistics and models. The LMGP-S's MSE is about 1\% of the QReg and 20\% of the MARS.
	
The functional prediction framework (implemented with LMGP-S here) can utilize more information in the data. As a result, it can achieve much better results for all summary statistics predictions. The traditional quantile regression does not have good predictions when dealing with complicated data with non-normal underlying distributions. Surrogates like MARS that use the scalar-form summary statistics directly also lose information, which shows the advantage of the proposed model framework.
	
	\begin{table}
		\centering
		\caption{MSEs for the predictions of multiple summary statistics. The QReg cannot predict the sample mean and standard deviation. MSEs are in the unit $10^{14}$  $(\text{KB/s})^2$.}\label{tab:SumStat}
\begin{center}
		\begin{tabular}{cccc}
			\hline
			& LMGP-S & QReg & MARS \\
			\hline
			Mean & 0.0012 & n/a & 0.0091 \\
			SD & 0.0013 & n/a & 0.0040 \\
			$Q( 0.05 )$ & 0.0033 & 0.3443 & 0.0065 \\
			$Q( 0.10 )$ & 0.0028 & 0.3830 & 0.0068 \\
			$Q( 0.25 )$ & 0.0019 & 0.4565 & 0.0067 \\
			$Q( 0.50 )$ & 0.0109 & 0.5727 & 0.0168 \\
			$Q( 0.75 )$ & 0.0030 & 0.4220 & 0.0238 \\
			$Q( 0.90 )$ & 0.0046 & 0.4325 & 0.0365 \\
			$Q( 0.95 )$ & 0.0057 & 0.4554 & 0.0420 \\
			\hline
		\end{tabular}
\end{center}
	\end{table}
	\section{Conclusions and Areas for Future Research}\label{sec:dis}
	In this paper, we focus on using the spline representation and Gaussian process to predict the I/O throughput distributions given the HPC system configuration. I-splines are used to represent the quantile function and the SVD is used to reduce the dimension. GP-based models are used to predict the SVD scores. The two LMGP models can be viewed as a mixture of the GP and CGP, and they can determine the proportion of GP and CGP automatically. We conduct comparisons between our model framework with some baseline methods. Numerical results show that our prediction framework has good performance in predictions for different subsets of the IOzone data, both in distributional and summary-statistic levels.
	
	One important future step in the management of performance variability is to develop a general tool to predict the throughput distribution for a new system configuration. Our LMGP models capture the relation between the system configuration and throughput distribution. One direct engineering application of the LMGP models is that one can utilize the predicted summary statistics to optimize the HPC system for different perspectives. For example, if we want to ensure a lower bound of the throughput, the 0.2 quantile can be part of the optimization objective. The advantage of our models is that LMGP models can use the distributional information and conduct much more accurate predictions on summary statistics.
	
	When comparing the discrepancy between two distributions, the Kolmogorov–Smirnov (KS) distance is used in some applications. We did not use the KS distance because, in certain situations, the KS distance can be misleading and is sensitive to a distributional shift. In particular, the KS distance can be misleading when a CDF $F(x)$ has a steep behavior (i.e., throughputs have multiple modes), and \shortciteN{xu2020modeling} show that multimodal behaviors commonly exist through the IOzone data. The KS distance measures the maximal error while $EL_1$ provides an average discrepancy. We use $EL_1$ as the error measurements in this paper, which is more appropriate.
	
One future research area is on data collection. As the number of factors becomes large, it will become impractical to collect a dataset as ``dense'' as in Figure~\ref{fig:fda_inidist} with functional responses. In this case, it will be interesting to explore a more sparse design in higher dimension as a ``screening'' step (e.g., \citeNP{DeanLewis2006}) to determine which system parameters are most critical for prediction, followed by more extensive data collection in the corresponding subspace.

Restricting Gaussian process models is not a simple matter (e.g., \citeNP{MitchellMorris1992}). The approach we used by projecting those negative weights back to the constrained space provides a convenient solution and the results are reasonably well. In the future, it will be interesting to explore other methods in constraining Gaussian process models, such as those described in \shortciteN{Swileretal2020}.

Note that the number of parameters for $\sigmamat_{\alphavec}$ increases at the order of $O(c^2)$. So when we have many categorical levels, the optimization for $\loglik_{2}$ will be difficult. As a result, a better parametrization for the categorical inputs can be investigated in future research. In addition, the EM algorithm is computationally intensive when we have a large dataset. Inversion of $\sigmamat_{\alphavec}$ is expensive because it becomes a dense matrix when $r_{\text{max}}$ is large. In the future, the estimation efficiency for using a sparse $\sigmamat_{\alphavec}$ can be studied.

Another future research is using positive matrix factorizations instead of SVD when decorrelating $\Bmat$. Currently, we use SVD and will introduce negative entries in $\wmat$. Several methods for positive matrix factorizations in \shortciteN{hopke2000guide} can be studied for decorrelating $\Bmat$. We will also perform simulation studies to further study the model estimability to predict $\wvec$. Last but not least, it will be interesting to study how the errors are propagated from I-spline smoothing to LMGP modeling in our prediction framework.

\section*{Acknowledgments}\label{Acknowledgement}
The authors acknowledge Advanced Research Computing at Virginia Tech for providing computational resources. The research was supported by National Science Foundation Grants CNS-1565314 and CNS-1838271 to Virginia Tech.

	\appendix
	\section{Formulas for the $\mathcal{Q}$ Functions}\label{sec:ch4ap1}
	We show the formulas for the
	$\mathcal{Q}_1\left[\thetavec_{\epsilonvec}|\thetavec_{\epsilonvec}^{(t-1)}\right]$ and $\mathcal{Q}_2\left[\thetavec_{\alphavec}|\thetavec_{\alphavec}^{(t-1)}\right]$. After the conditional expectation is taken, we obtain,
	\begin{align*}
		&\mathcal{Q}_1\left[\thetavec_{\epsilonvec}|\thetavec_{\epsilonvec}^{(t-1)}\right]=\E_{\alphavec|\wvec,\thetavec_{\vepsvec}^{(t-1)},\thetavec_{\alphavec}^{(t-1)}}\loglik_{1}(\thetavec_{\vepsvec}; \wvec|\alphavec)\\
		=&\E_{\alphavec|\wvec,\thetavec_{\vepsvec}^{(t-1)},\thetavec_{\alphavec}^{(t-1)}}\left[-\frac{n}{2}\log(2\pi)-\frac{1}{2}\log(|\sigmamat_{\vepsvec}|)-\frac{1}{2}
		(\wvec-\muvec-\alphavec)^{\T}\sigmamat_{\vepsvec}^{-1}(\wvec-\muvec-\alphavec)\right]\\
		=&-\frac{n}{2}\log(2\pi)-\frac{1}{2}\log(|\sigmamat_{\epsilonvec}|)-\frac{1}{2}(\wvec-\muvec)^{\T}\sigmamat_{\vepsvec}^{-1}(\wvec-\muvec)\\
		&+\frac{1}{2}\E_{\alphavec|\wvec,\thetavec_{\vepsvec}^{(t-1)},\thetavec_{\alphavec}^{(t-1)}}\left[2\alphavec^{\T}\sigmamat_{\vepsvec}^{-1}(\wvec-\muvec)-\alphavec^{\T}\sigmamat_{\epsilonvec}\alphavec\right]\\
		=&-\frac{n}{2}\log(2\pi)-\frac{1}{2}\log(|\sigmamat_{\epsilonvec}|)-\frac{1}{2}(\wvec-\muvec)^{\T}\sigmamat_{\vepsvec}^{-1}(\wvec-\muvec)\\
		&+(\wvec-\muvec)^{\T}\sigmamat_{\epsilonvec}^{-1}\muvec_{\alphavec|\wvec}-\frac{1}{2}\tr\left[\sigmamat_{\epsilonvec}^{-1}\sigmamat_{\alphavec|\wvec}^{(t-1)}\right]-\frac{1}{2}\left[\muvec_{\alphavec|\wvec}^{(t-1)}\right]^{\T}\sigmamat_{\epsilonvec}^{-1}\muvec_{\alphavec|\wvec}^{(t-1)},\\
		\text{and }&&\\
		&\mathcal{Q}_2\left[\thetavec_{\alphavec}|\thetavec_{\alphavec}^{(t-1)}\right]=\E_{\alphavec|\wvec,\thetavec_{\vepsvec}^{(t-1)},\thetavec_{\alphavec}^{(t-1)}}\loglik_{2}(\thetavec_{\alphavec};\alphavec)\\
		=&\E_{\alphavec|\wvec,\thetavec_{\vepsvec}^{(t-1)},\thetavec_{\alphavec}^{(t-1)}}\left[-\frac{n}{2}\log(2\pi)-\frac{1}{2}\log(|\sigmamat_{\alphavec}|)-\frac{1}{2}
		\alphavec^{\T}\sigmamat_{\alphavec}^{-1}\alphavec\right]\\
		=&-\frac{n}{2}\log(2\pi)-\frac{1}{2}\log(|\sigmamat_{\alphavec}|)-\frac{1}{2}\tr\left[\sigmamat_{\alphavec}^{-1}\sigmamat_{\alphavec|\wvec}^{(t-1)}\right]-\frac{1}{2}\left[\muvec_{\alphavec|\wvec}^{(t-1)}\right]^{\T}\sigmamat_{\alphavec}^{-1}\muvec_{\alphavec|\wvec}^{(t-1)}.
	\end{align*}
	
	\section{Derivations for Different Parameters in Each Category}\label{sec:ch4ap2}
	When we have different $\mu,\nuvec,g$, and $\sigma_{\epsilonvec}^2$ for each category, the $\thetavec_{\epsilonvec}$ becomes
$$\thetavec_{\epsilonvec}=(\sigma_{\epsilonvec,1}^2,\mu_1,\nuvec_1^\T,g_1,\ldots,\sigma_{\epsilonvec,k}^2,\mu_k,\nuvec_k^\T,g_k, \ldots,\sigma_{\epsilonvec,c}^2,\mu_c,\nuvec_c^\T,g_c)^\T.$$
	Let $\thetavec_{\epsilonvec,k}=(\sigma_{\epsilonvec,k}^2,\mu_k,\nuvec_k^\T,g_k)^\T$, the $\sigmamat_{\vepsvec}$ becomes
	\begin{align*}
		\sigmamat_{\vepsvec}=
\text{Diag}
		\left(
			\sigmamat_{\vepsvec,\mathbb{I}_1},\ldots,\sigmamat_{\vepsvec,\mathbb{I}_k},\ldots, \sigmamat_{\vepsvec,\mathbb{I}_c}
		\right)=
\text{Diag}
		\left(
			\sigma_{\vepsvec,1}^2\omegamat_{\vepsvec,\mathbb{I}_1},\ldots,\sigma_{\vepsvec,k}^2\omegamat_{\vepsvec,\mathbb{I}_k}, \ldots, \sigma_{\vepsvec,c}^2\omegamat_{\vepsvec,\mathbb{I}_c}
		\right).
	\end{align*}
	Using the block diagonal structure of $\sigmamat_{\epsilonvec}$, the $\mathcal{Q}_1\left[\thetavec_{\epsilonvec}|\thetavec_{\epsilonvec}^{(t-1)}\right]$ in (\ref{eq:q1}) becomes:
	\begin{align*}
		&\mathcal{Q}_1\left[\thetavec_{\epsilonvec}|\thetavec_{\epsilonvec}^{(t-1)}\right]=
		-\frac{n}{2}\log(2\pi)-\frac{1}{2}\sum_{k=1}^{c}\log(|\sigmamat_{\epsilonvec,\mathbb{I}_k}|)-\frac{1}{2}\sum_{k=1}^{c}(\wvec_{\mathbb{I}_k}-\muvec_{\mathbb{I}_k})^{\T}\sigmamat_{\vepsvec,\mathbb{I}_k}^{-1}(\wvec_{\mathbb{I}_k}-\muvec_{\mathbb{I}_k})\\
		&+\sum_{k=1}^{c}(\wvec_{\mathbb{I}_k}-\muvec_{\mathbb{I}_k})^{\T}\sigmamat_{\epsilonvec,\mathbb{I}_k}^{-1}\muvec_{\alphavec|\wvec,\mathbb{I}_k}^{(t-1)}-\frac{1}{2}\sum_{k=1}^{c}\tr\left[\sigmamat_{\epsilonvec,\mathbb{I}_k}^{-1}\sigmamat_{\alphavec|\wvec,\mathbb{I}_k}^{(t-1)}\right]-\frac{1}{2}\sum_{k=1}^{c}\left[\muvec_{\alphavec|\wvec,\mathbb{I}_k}^{(t-1)}\right]^{\T}\sigmamat_{\epsilonvec,\mathbb{I}_k}^{-1}\muvec_{\alphavec|\wvec,\mathbb{I}_k}^{(t-1)}\\
		=&\sum_{k=1}^{c}\mathcal{Q}_1\left[\thetavec_{\epsilonvec,\mathbb{I}_k}|\thetavec_{\epsilonvec,\mathbb{I}_k}^{(t-1)}\right],\\
		\text{where, }&&\\
		&\mathcal{Q}_1\left[\thetavec_{\epsilonvec,\mathbb{I}_k}|\thetavec_{\epsilonvec,\mathbb{I}_k}^{(t-1)}\right]=-\frac{n_k}{2}\log(2\pi)-\frac{1}{2}\log(|\sigmamat_{\epsilonvec,\mathbb{I}_k}|)-\frac{1}{2}(\wvec_{\mathbb{I}_k}-\muvec_{\mathbb{I}_k})^{\T}\sigmamat_{\vepsvec,\mathbb{I}_k}^{-1}(\wvec_{\mathbb{I}_k}-\muvec_{\mathbb{I}_k})\\
		&+(\wvec_{\mathbb{I}_k}-\muvec_{\mathbb{I}_k})^{\T}\sigmamat_{\epsilonvec,\mathbb{I}_k}^{-1}\muvec_{\alphavec|\wvec,\mathbb{I}_k}^{(t-1)}-\frac{1}{2}\tr\left[\sigmamat_{\epsilonvec,\mathbb{I}_k}^{-1}\sigmamat_{\alphavec|\wvec,\mathbb{I}_k}^{(t-1)}\right]-\frac{1}{2}\left[\muvec_{\alphavec|\wvec,\mathbb{I}_k}^{(t-1)}\right]^{\T}\sigmamat_{\epsilonvec,\mathbb{I}_k}^{-1}\muvec_{\alphavec|\wvec,\mathbb{I}_k}^{(t-1)}.
	\end{align*}
	Thus, $\thetavec_{\epsilonvec,\mathbb{I}_k}$ can be estimated through maximizing $\mathcal{Q}_1\left[\thetavec_{\epsilonvec,\mathbb{I}_k}|\thetavec_{\epsilonvec,\mathbb{I}_k}^{(t-1)}\right]$ separately.
	\section{Formulas for the Score Functions}\label{sec:formula.score.function}
	We use the gradient-based method for optimization. Here we provide the approximated score function for the likelihoods $\mathcal{Q}_1$ and $\mathcal{Q}_2$ for the LMGP model.
	For $\mathcal{Q}_1$, we assume $\muhat$ does not depend on $\nuvec$ and $g$. Then, for $\nu_l$ ,$l=1,\ldots,p$, we have
	\begin{align*}
		\frac{\partial \mathcal{Q}_1}{\partial \nuvec}&=-\frac{1}{2}\text{tr}\left(\omegamat_{\epsilonvec}^{-1}\frac{\partial\omegamat_{\epsilonvec}}{\partial \nuvec}\right)-\frac{n}{2\sigmahat_{\epsilonvec}^2}\frac{\partial\sigmahat_{\epsilonvec}^2}{\partial\nuvec},\\
		\frac{\partial\left(\omegamat_{\epsilonvec}\right)_{ij}}{\partial \nu_l}&=-\frac{(x_{id}-x_{jd})^2}{\nu_l^2}\exp\left[-\sum_{l=1}^p\frac{(x_{il}-x_{jl})^2}{\nu_l}\right],\\
		n\frac{\partial\sigmahat_{\epsilonvec}^2}{\partial\nu_l}&=(\wvec-\wh\muvec)^{\T}\omegamat_{\vepsvec}^{-1}\frac{\partial \omegamat_{\epsilonvec}}{\partial\nu_l}\omegamat_{\vepsvec}^{-1}(\wvec-\wh\muvec)-
		2(\wvec-\wh\muvec)^{\T}\omegamat_{\vepsvec}^{-1}\frac{\partial \omegamat_{\epsilonvec}}{\partial\nu_l}\omegamat_{\vepsvec}^{-1}\muvec_{\alphavec|\wvec}\\
		&+\tr\left[\omegamat_{\vepsvec}^{-1}\frac{\partial \omegamat_{\epsilonvec}}{\partial\nu_l}\omegamat_{\vepsvec}^{-1}\sigmamat_{\alphavec|\wvec}^{(t-1)}\right]+\left[\muvec_{\alphavec|\wvec}^{(t-1)}\right]^{\T}\omegamat_{\vepsvec}^{-1}\frac{\partial \omegamat_{\epsilonvec}}{\partial\nu_l}\omegamat_{\vepsvec}^{-1}\muvec_{\alphavec|\wvec}^{(t-1)}.
	\end{align*}
	For $g$, we can get $\frac{\partial \mathcal{Q}_1}{\partial g}$ similarly using $\frac{\partial\omegamat_{\epsilonvec}}{\partial g}=\mathbf{I}$.
	
	For $\mathcal{Q}_2$, we first have an alternative expression for $\omegamat_{\alphavec}$,
	$
		\omegamat_{\alphavec}=\Amat^\T\left(\Pmat\otimes\mathbf{\Phi}\right)\Amat,
	$
	where $\otimes$ is the Kronecker product, $\mathbf{\Phi}$ is an $n\times n$ matrix satisfies
	$
		{\Phi}_{ii'}=\kappa(r_{ii'},r_{\text{max}}),
	$
	and $\Amat$ is an $n\times c$ matrix. The $i$th row of $\Amat$ has the $k$th element equal to 1 and the rest $(c-1)$ elements equal to 0, where $k$ is the level of $\zvec_i$ after sorting. Then we have
	\begin{align*}
		\frac{\partial \mathcal{Q}_2}{\partial \rhovec}&=-\frac{1}{2}\text{tr}\left(\omegamat_{\alphavec}^{-1}\frac{\partial\omegamat_{\alphavec}}{\partial \rhovec}\right)-\frac{n}{2\sigmahat_{\alphavec}^2}
\frac{\partial\sigmahat_{\alphavec}^2}{\partial\rhovec},\quad\quad
		\frac{\partial\omegamat_{\alphavec}}{\partial \rhovec}=\Amat^\T\left(\frac{\partial\Pmat}{\partial\rhovec}\otimes\mathbf{\Phi}\right)\Amat,\\ n\frac{\partial\sigmahat_{\alphavec}^2}{\partial\rhovec}&=\tr\left[\omegamat_{\alphavec}^{-1}\frac{\partial\omegamat_{\alphavec}}{\partial \rhovec}\omegamat_{\alphavec}^{-1}\sigmamat_{\alphavec|\wvec}^{(t-1)}\right]+\left[\muvec_{\alphavec|\wvec}^{(t-1)}\right]^{\T}\omegamat_{\alphavec}^{-1}\frac{\partial\omegamat_{\alphavec}}{\partial \rhovec}\omegamat_{\alphavec}^{-1}\muvec_{\alphavec|\wvec}^{(t-1)}.
	\end{align*}
	Then ${\partial \omegamat_{\alphavec}}/{\partial \rhovec}$ can be expressed using ${\partial\Pmat}/{\partial \rhovec}$.


\end{document}